\documentclass[aps,prx,showpacs,citeautoscript,superscriptaddress,longbibliography,notitlepage,twocolumn]{revtex4-1}

\usepackage{graphicx}  
\usepackage{epstopdf}
\usepackage{dcolumn}   
\usepackage{bm}        
\usepackage{amsmath,amssymb,amsfonts,bm,dsfont,units}
\usepackage{wrapfig}
\usepackage{array}
\usepackage[caption=false]{subfig}
\usepackage[bookmarks=false,linkcolor=blue,urlcolor=blue,colorlinks,citecolor=blue]{hyperref}
\usepackage{exscale,relsize}
\usepackage{verbatim}
\usepackage[usenames, dvipsnames]{color}
\newcommand{\ket}[1]{\left|#1\right>}
\newcommand{\bra}[1]{\left<#1\right|}
\usepackage{hyperref}
\usepackage{braket}

\begin{document}

	\global\long\def\dg{^{\dagger}}
	\global\long\def\sm{\sum_{<i,j>}}
	\global\long\def\up{\uparrow}
	\global\long\def\dn{\downarrow}
	\global\long\def\ri{\mathbf{r_{i}}}
	\global\long\def\rj{\mathbf{r_{j}}}
	\global\long\def\sg{\sigma}
	\global\long\def\si{\mathbf{S_{i}}}
	\global\long\def\sj{\mathbf{S_{j}}}
	\global\long\def\pp{\mathcal{P}_{0}}
	\global\long\def\qq{\mathcal{Q}_{0}}
	\global\long\def\qi{\frac{1}{\qq H_{0}\qq-E_{i}}}
	\global\long\def\qj{\frac{1}{\qq H_{0}\qq-E_{j}}}
	\global\long\def\qk{\frac{1}{\qq H_{0}\qq-E_{k}}}
	\global\long\def\jo{J_{n}(\zeta_{12})}
	\global\long\def\ji{J_{n+1}(\zeta_{12})}
	\global\long\def\js{J_{n+2}(\zeta_{12})}
	\global\long\def\nu{\Ket{0}}
\title{Controlling ligand-mediated exchange interactions in periodically driven magnetic materials}	
\author{Swati Chaudhary}
\affiliation{Department of Physics and Institute for Quantum Information and Matter, California Institute of Technology, Pasadena, California 91125, USA}
\author{Alon Ron}
\affiliation{Raymond and Beverly Sackler School of Physics and Astronomy, Tel-Aviv University, Tel Aviv, 69978, Israel}
\author{David Hsieh}
\affiliation{Department of Physics and Institute for Quantum Information and Matter, California Institute of Technology, Pasadena, California 91125, USA}
\author{Gil Refael}
\affiliation{Department of Physics and Institute for Quantum Information and Matter, California Institute of Technology, Pasadena, California 91125, USA}

\begin{abstract}
A periodic drive could alter the effective exchange interactions in magnetic materials. Here,
we explore how exchange pathways affect the effective interactions of periodically driven magnetic materials. Aiming to apply Floquet engineering methods to two-dimensional magnetic materials, we consider realistic models and discuss the effect of a periodic drive on ligand-mediated exchange interactions. We show that depending on bond angles and the number of ligand ions involved in the exchange process, drive-induced changes can be very different from those calculated from direct-hopping models considered earlier. We study these effects and find that the presence of ligand ions must be taken into account, especially for TMTCs where ligand ion mediated next-neighbor interactions play a crucial role in determining the magnetic ground state of the system.
\end{abstract}
\maketitle	
\section{Introduction}

Periodic drives have been used extensively to tailor the properties of the hamiltonians for ultracold gases in optical lattices, ranging from the generation of artificial gauge fields for neutral atoms in optical lattices to many body localization \cite{F_PhysRevB.79.081406,F_PhysRevB.84.235108,F_PhysRevLett.105.017401,F_lindner2011floquet,F_PhysRevX.4.031027,F_PhysRevX.6.021013,F_PhysRevLett.110.026603,F2_PhysRevB.96.020201,F_PhysRevLett.110.026603,F_fausti2011light,F_PhysRevLett.116.176401,F_PhysRevLett.112.156801,F2_PhysRevB.96.205127,F2_PhysRevLett.109.145301,F2_PhysRevLett.118.115301,F2_baum2018dynamical,F2_goldman2014light,F2_holthaus2015floquet,F2_itin2015effective,FK_PhysRevB.97.085405,FK_PhysRevB.95.104308,F_PhysRevB.95.045102,FE_PhysRevB.96.155435,FM_PhysRevB.95.134508,FM_PhysRevB.95.155407,FM_PhysRevB.90.205127,FM_PhysRevB.96.155438,FM_yang2018floquet,FM_PhysRevB.95.174306,FSL_jotzu2014experimental,FQ_PhysRevB.97.035123,MagFl_owerre2018photoinduced,F0_PhysRevB.96.205127,F0_PhysRevB.96.104309,F0_PhysRevB.97.075420,F0_PhysRevLett.119.267701,F0_inoue2018floquet,FKondo_PhysRevB.96.115120,Fweyl_PhysRevB.94.235137,FWeyl_PhysRevB.94.121106,FWeyl_fu2017phase,Fweyl_niu2018tunable,Fweyl_PhysRevB.96.041205,FWeyl_PhysRevB.96.041206,Fweyl_PhysRevB.96.041126,Fweyl_PhysRevB.95.144311,Fweyl2_PhysRevB.96.041205,Fweyl_PhysRevLett.117.087402}. The evolution of such periodically-driven systems can be described by an effective time-independent hamiltonian using  Floquet theory~\cite{PhysRev.138.B979}. The properties of this effective hamiltonian can be controlled by changing the drive parameters like its frequency, amplitude etc. 
Extending these methods to quantum materials seems very promising as it may allow us to realize new states of matter and manipulate the electronic and magnetic properties of these materials on demand ~\cite{basov2017towards,oka2018floquet}.

 Previously, several works studied light-induced changes in the magnetic properties of transition metals compounds~\cite{Pkirilyuk2010ultrafast,mikhaylovskiy2015ultrafast,mentink2015ultrafast,Mentink2014,mentink2015ultrafast,mentink2017manipulating,Bukov2016,hejazi1PhysRevLett.121.107201,hejazi2018floquet,quito2020floquet,quito2020polarization,PhysRevB.100.220403,ron2019ultrafast}.  
 Recent works ~\cite{Mentink2014,mentink2015ultrafast,mentink2017manipulating,Bukov2016,hejazi1PhysRevLett.121.107201,hejazi2018floquet,quito2020floquet,quito2020polarization,PhysRevB.100.220403} have demonstrated the possibility of using periodic drives for manipulating the exchange interactions in extended antiferromagnetic (AFM) Mott insulators. These results can be applied to many transition metal (TM) compounds. Transition metal trichalcogenide (TMTC) monolayers are one of the prime candidates, where periodic drive could lead to interesting results.  As shown in Ref.~\cite{Sivadas2015}, the magnetic properties of such monolayers  are very well described by the Heisenberg model on a honeycomb lattice with up to third nearest-neighbor interactions. Motivated by these materials, we study the effects of the periodic drive on a Fermi-Hubbard Model(FHM) on a honeycomb lattice, and study how the magnetic coupling strength can be modified by tuning  different drive parameters. We further explore the consequences  of the ligand ions and study how the changes in magnetic coupling strength depend not only on the drive parameters, but also on the bond angles and the orbital orientation of these intermediary ions.

\section{Periodically Driven Fermi Hubbard Model}
\label{FHM_honeycomb}
We study the effect of a periodic drive on the exchange interactions using a periodically driven Fermi Hubbard model (FHM) in the Mott regime at half-filling. Let us first review this model.
\subsection{Review : Toy Model for AFM coupling renormalization due to photo-modified tunneling }
 In the presence of a time-dependent electric field, the full Hamiltonian of the Fermi-Hubbard model is given by:
\begin{equation}
H=-t\sm c_{i\sigma}\dg c_{j\sigma}+\text{h.c}+U\sum_{i}n_{i\uparrow}n_{i\downarrow}+e\mathbf{E}\cdot\mathbf{\sum_{i,\sigma}}n_{i\sigma}\mathbf{r_{j}\cos(\omega t)}.
\end{equation}
After Peierls substitution, it becomes:
\begin{equation}
H'=-t\sm e^{i\left[\frac{ e\mathbf{E}\cdot(\mathbf{r_{j}-}\mathbf{r}_{i})}{\omega}\sin(\omega t)\right]}c_{i\sigma}\dg c_{j\sigma}+\text{h.c}+H_{U}=H_{t}'+H_{U}.
\end{equation}
In the limit $U\gg t$, and for a non-resonant drive, the exchange coupling is given by:
\begin{equation}
J_{i}'=J_{i}U\sum_{n=-\infty}^{\infty} \frac{1}{U+n\omega}\mathcal{J}_n(\zeta_i)^2,
\label{hoppingrenormalization}
\end{equation}
where, $J_i=\frac{4t^2}{U}$ is the magnetic coupling strength for the undriven case,  $\mathcal{J}_n$ denotes $n^{th}$ order Bessel function, and drive parameter
\begin{equation}
\zeta_i=\frac{e\textbf{E}\cdot\Delta \textbf{r}_i}{\omega},
 \end{equation}
 where $\Delta \textbf{r}_i$ is the displacement between $i^{th}$ neighbors.
  In the presence of this periodic drive, spin exchange interactions are affected mainly due to two factors: (a) change in the hopping parameter due to photon-assisted tunneling, and  (b) virtual excitations between different Floquet sectors. The effective spin exchange interactions can thus be controlled by changing the frequency, polarization or intensity of the laser. Previous works~\cite{Mentink2014,mentink2015ultrafast,mentink2017manipulating,Bukov2016,hejazi1PhysRevLett.121.107201,hejazi2018floquet} have studied the periodically driven FHM extensively for both near-resonant and off-resonant cases. The above expression in Eq.~(\ref{hoppingrenormalization}) is valid only for a non-resonant drive where doublon sectors are well separated in energy from the single occupation sector. Near-resonant drive can be handled using a somewhat similar machinery of Floquet formalism as shown in Ref.~\cite{Bukov2016} but in certain cases, real doublon-holon pairs can significantly affect the exchange interactions~\cite{Liu2018}.

\begin{figure}
	\includegraphics[scale=0.48]{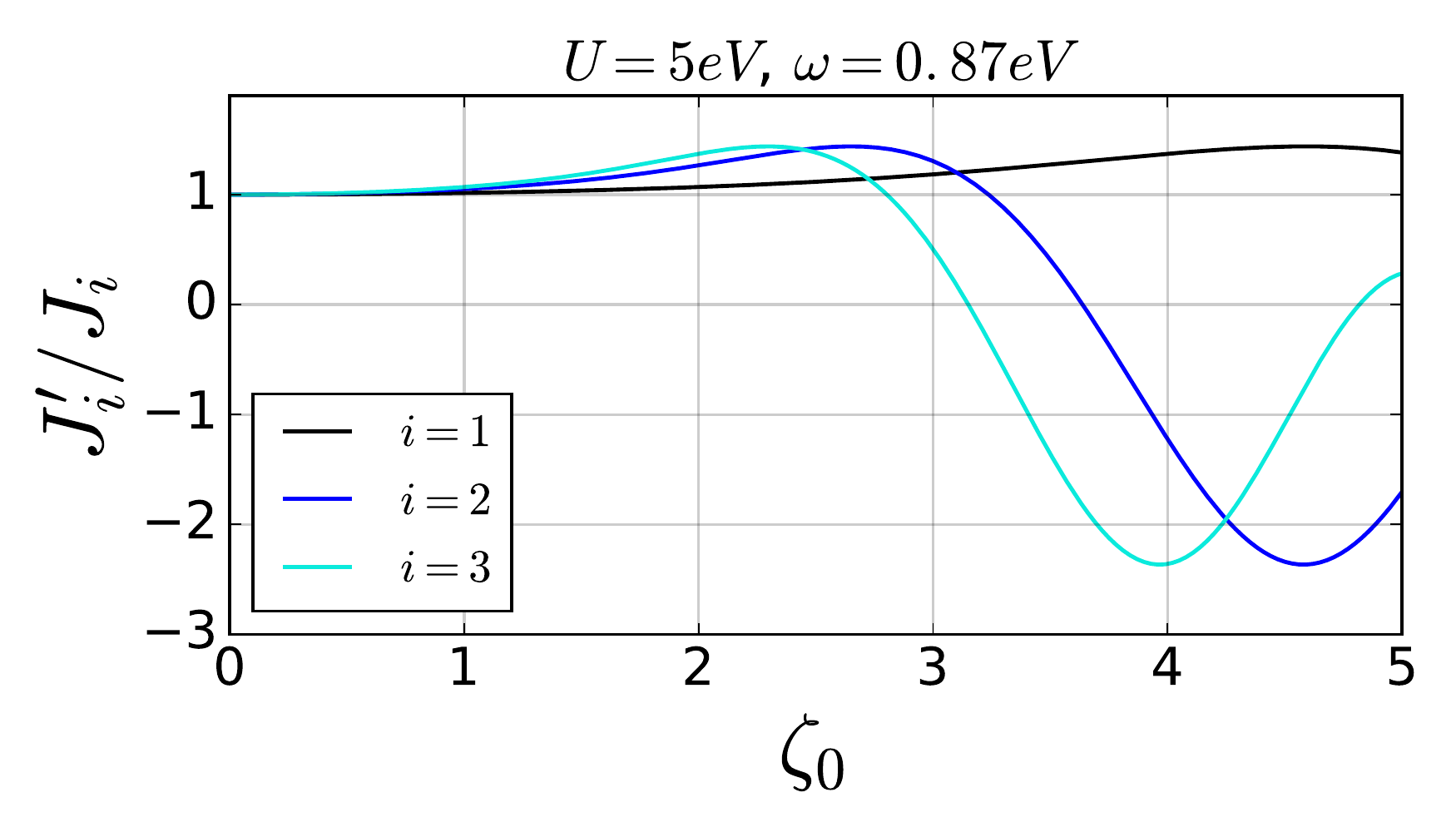}
	\includegraphics[scale=0.48]{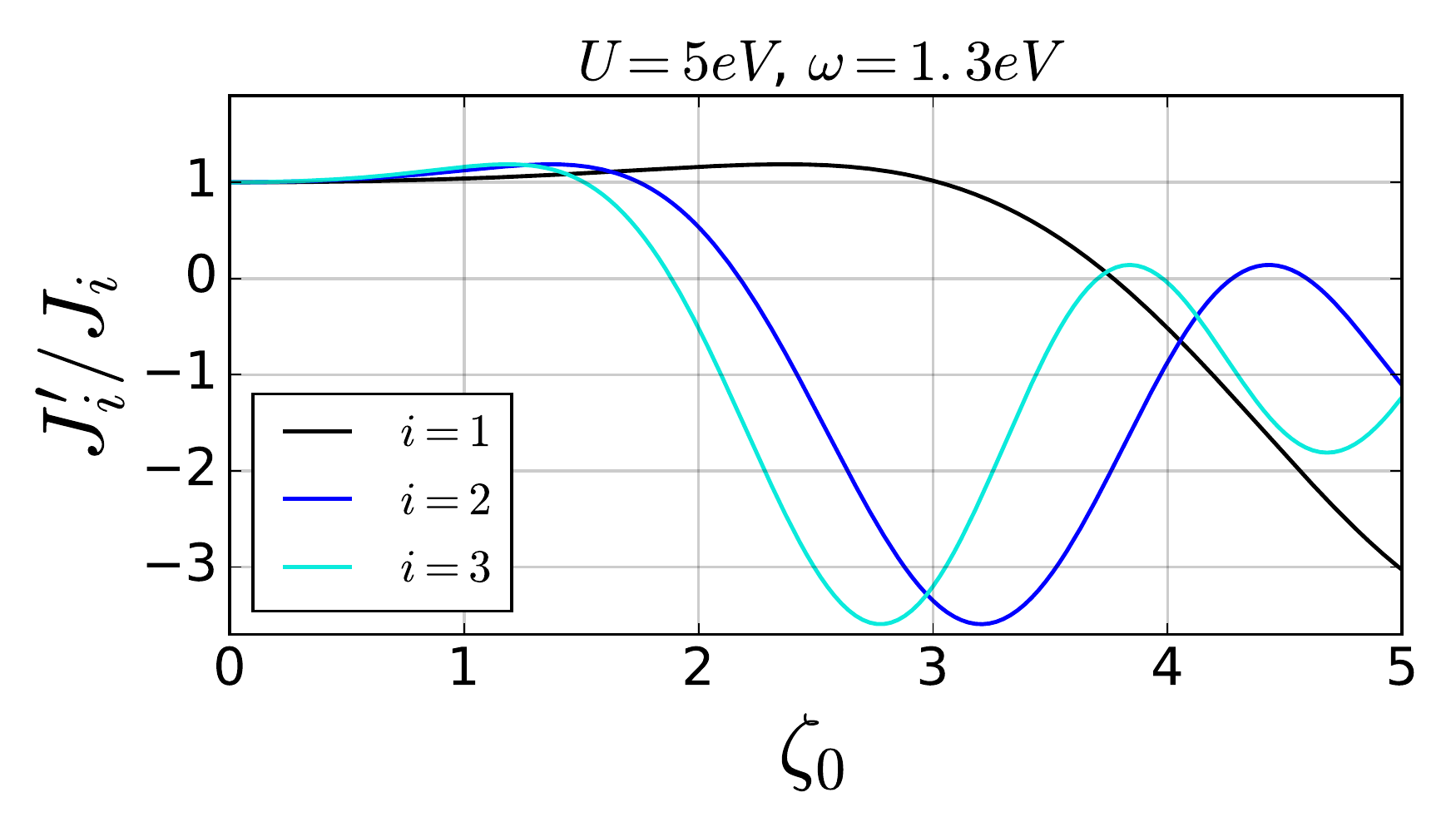}
	\caption{\textbf{Periodic drive effect on magnetic coupling.}  Changes in the spin exchange interaction energy for different neighbors as a function of the drive parameter $\zeta_0$ (in terms of E field magnitude, one unit on this scale can be read as $1V/\AA$) for two different values of $U/\omega$. The changes are larger for smaller values of $U/\omega$ as expected from Eq. \ref{hoppingrenormalization}.}
	\label{renormcoupling}
\end{figure}

\subsection{Driven FHM on Honeycomb lattice }
We are interested in controlling the properties of magnetic materials using light, and monolayer magnetic materials, e.g TMTC monolayers, provide a suitable platform for our exploration. The magnetic structure of these 2D magnetic materials is captured by the Heisenberg model on a honeycomb lattice with up to third nearest neighbor interactions. It exhibits numerous ground states depending on the relative signs and values of different neighbor exchange interactions~\cite{Sivadas2015}.  So, before proceeding further, we briefly consider the effects of a non-resonant periodic drive on the exchange interactions in this model.

\begin{figure}
	\centering

	\includegraphics[scale=0.35]{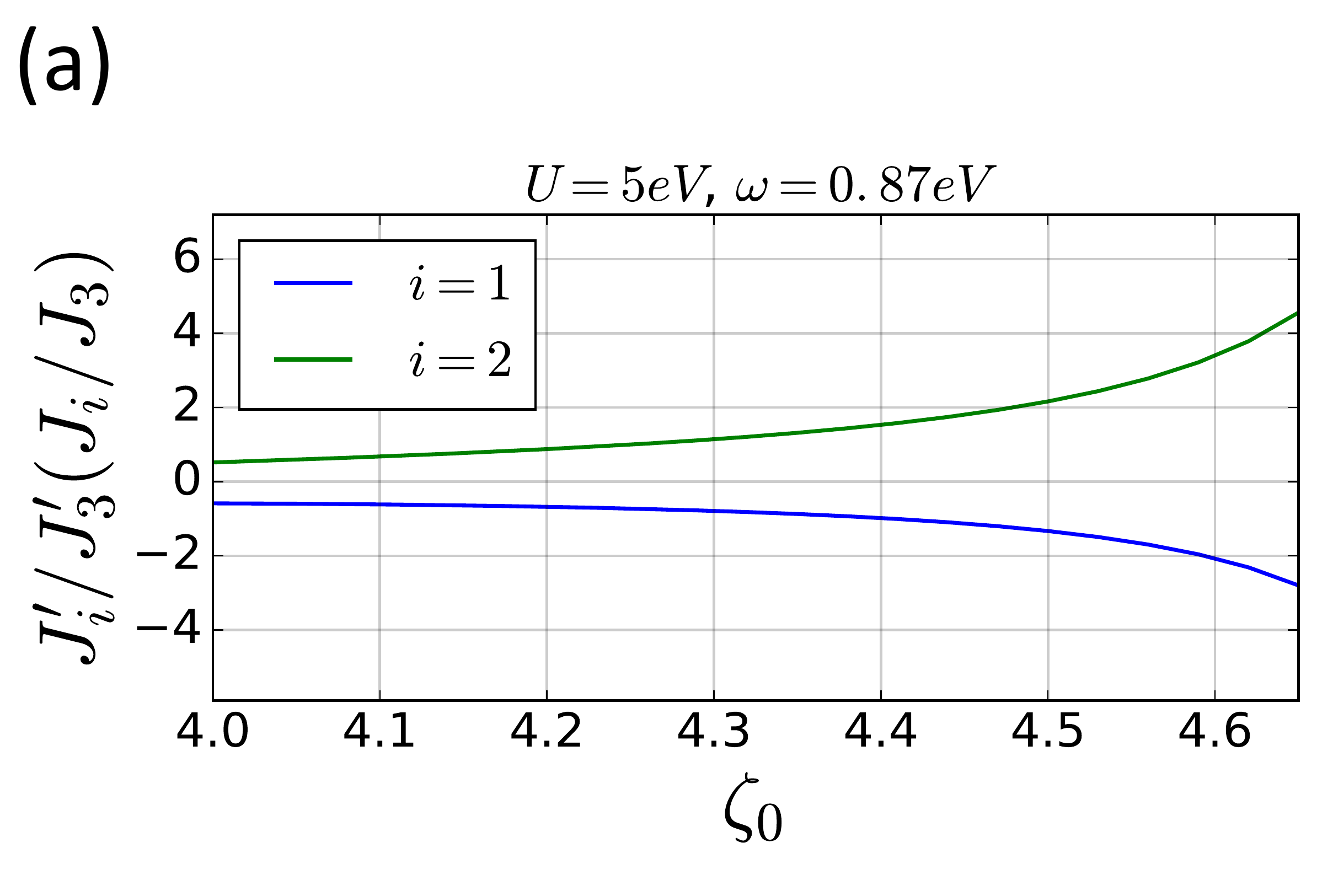}
	\includegraphics[scale=0.35]{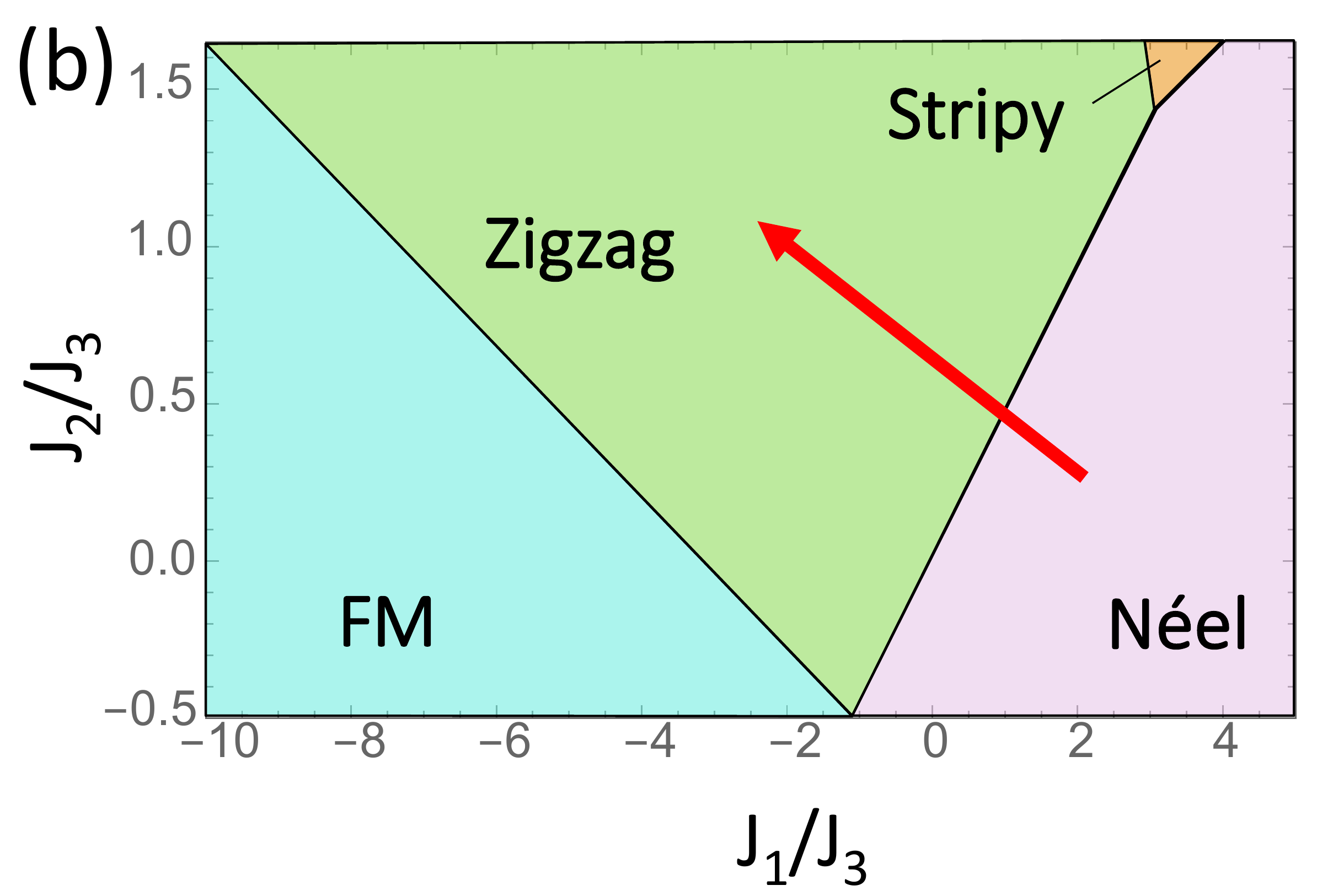}
	\caption{\textbf{Magnetic ground state of effective Hamiltonian.} (a) Change in the magnetic coupling strength ratio as a function of the drive parameter, and (b) same change shown by a red arrow on the phase diagram \cite{Sivadas2015}.}
	\label{ratiochange}
\end{figure}

Consider the Fermi-Hubbard Model on a honeycomb lattice with up to third nearest neighbor hopping:
\newcommand{\jj}{\mathcal{J}}
\begin{equation}
\begin{split}
H=&H_{U}+H_{t}=U\sum_{i}\hat{n}_{i\uparrow}\hat{n}_{i\downarrow}+\sum_{\left<i,j\right>\sg}t_{1}c_{\ri\sg}\dg c_{\rj\sg}\\&+\sum_{\left<\left<i,j,\sg\right>\right>}t_{2}c_{\ri\sg}\dg c_{\rj\sg}+\sum_{\left<\left<\left<i,j\right>\right>\right>}t_{3}c\dg_{\ri\sg}c_{\rj\sg}+\text{h.c}
\end{split}
\end{equation}
where $U\gg t_i$. Adding a circularly polarized light beam gives rise to the following periodic drive:
\begin{equation}
H'(t)=\sum_ieE_0(\cos\omega t\,\hat{x} +\sin\omega t\,\hat{y})\cdot\textbf{r}_i \hat{n}_i
\end{equation}
which results in a direction-independent modification of exchange interactions.

In the presence of this drive, the magnetic coupling strength between $i^{\text{th}}$ neighbors is given by an expression similar to Eq.~(\ref{hoppingrenormalization}) with $t$ replaced by $t_i$, and the changes with drive parameter $\zeta_0=\frac{eE_0 a}{\omega}$, where $E_0$ is the electric field amplitude, $e$ is the electron charge and $\omega$ is the frequency of drive measured in units of eV, and $a$ is the separation between the nearest neighbors on a honeycomb lattice, are shown in~Fig.~\ref{renormcoupling}. This drive parameter is dimensionless.  In most of the cases, $a\approx1\AA$ and $\omega\approx 1eV$, and thus $\zeta=1$, roughly corresponds to an electric field amplitude of $1\text{V}/\AA$, i.e $100\text{MV/cm}$.
The changes in the coupling strength depend on both the drive parameter $\zeta_0$, and the ratio $U/\omega$. The main contribution in Eq.~(\ref{hoppingrenormalization}) comes from those values of $n$ which are close to $U/\omega$. Roughly speaking, this change in the coupling constant behaves in the same manner as $\mathcal{J}_n(\zeta_i)$, and thus for large values of $U/\omega$, the renormalization factor peaks at a higher value of the drive parameter, and its amplitude is also smaller as we notice in Fig.~\ref{renormcoupling}.
This model is particularly interesting because depending on the drive parameter, the ground state of the effective time-independent hamiltonian can be very different from the undriven case as shown in Fig.~\ref{ratiochange}.

\section{Applications to Materials with ligand mediated magnetic interactions}
\begin{figure*}
	\includegraphics[scale=0.5]{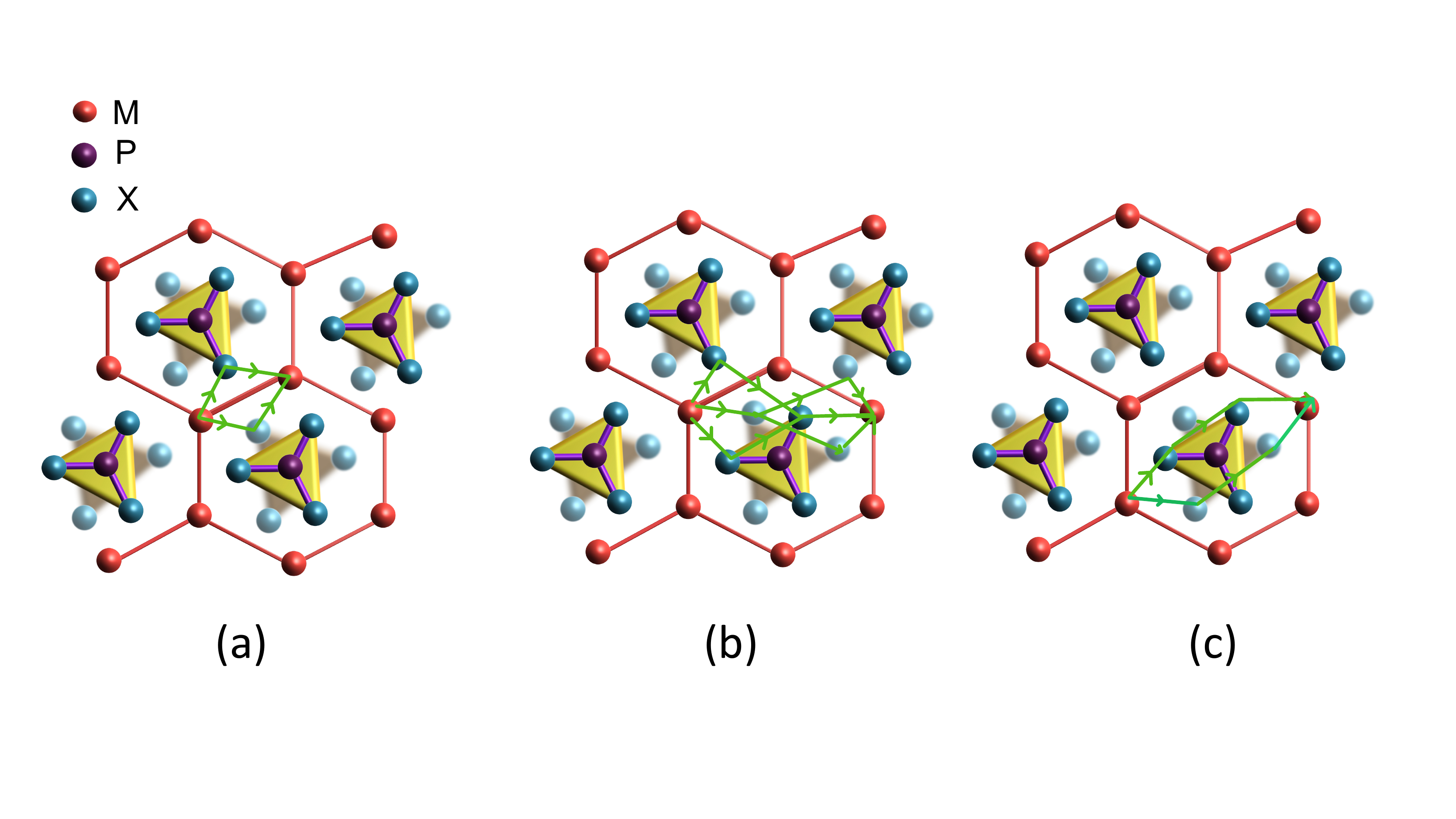}
	\caption{\textbf{Exchange Pathways in TMTC monolayers.} Top view of single layer metal phosphorus trichalcogenides ($\text{MPX}_3$)  illustrating different paths responsible for spin exchange interaction (a) Nearest neighbor interactions $J_1$ can occur via direct hopping or via one ligand ion (b) Second nearest neighbor interactions are mediated by two intermediate ions, and there are five paths available for spin exchange (c) Third nearest neighbor interactions occur via two intermediate ions and both belong to the same plane.}
	\label{exchangepathwayMnPX3}
	
\end{figure*}

Most of the previous works~\cite{Mentink2014,mentink2015ultrafast,mentink2017manipulating,Bukov2016,hejazi1PhysRevLett.121.107201,hejazi2018floquet} on the periodically driven FHM assumed direct hopping between two metal sites. In TM compounds spin-exchange interactions are mostly mediated by ligand ions as shown in Fig.~\ref{exchangepathwayMnPX3} for TMTC monolayers \cite{Sivadas2015}, and thus the exchange coupling depends on factors like bond lengths, bond angles, and the nature of orbitals involved in the exchange process. There are usually multiple  pathways available for spin-exchange processes between two metal sites. 
Particularly for TMTC monolayers~\cite{Sivadas2015}, the nearest neighbor interactions occur via direct hopping or via one ligand ion, second and third nearest neighbor interactions involve indirect hopping mediated by two ligand ions as shown in Fig.~\ref{exchangepathwayMnPX3}. In order to provide a  more precise estimate of the change in coupling strength, one must take these factors into account. Below, we explore the consequences of periodic drive for different cases, and contrast them with the periodic drive effects for the direct-hopping case.\\
\subsection*{Case 1 : AF coupling via one intermediate ion with only one orbital}
\label{superexchangeAF}

\begin{figure}[h]
	\centering
	\includegraphics[scale=0.4]{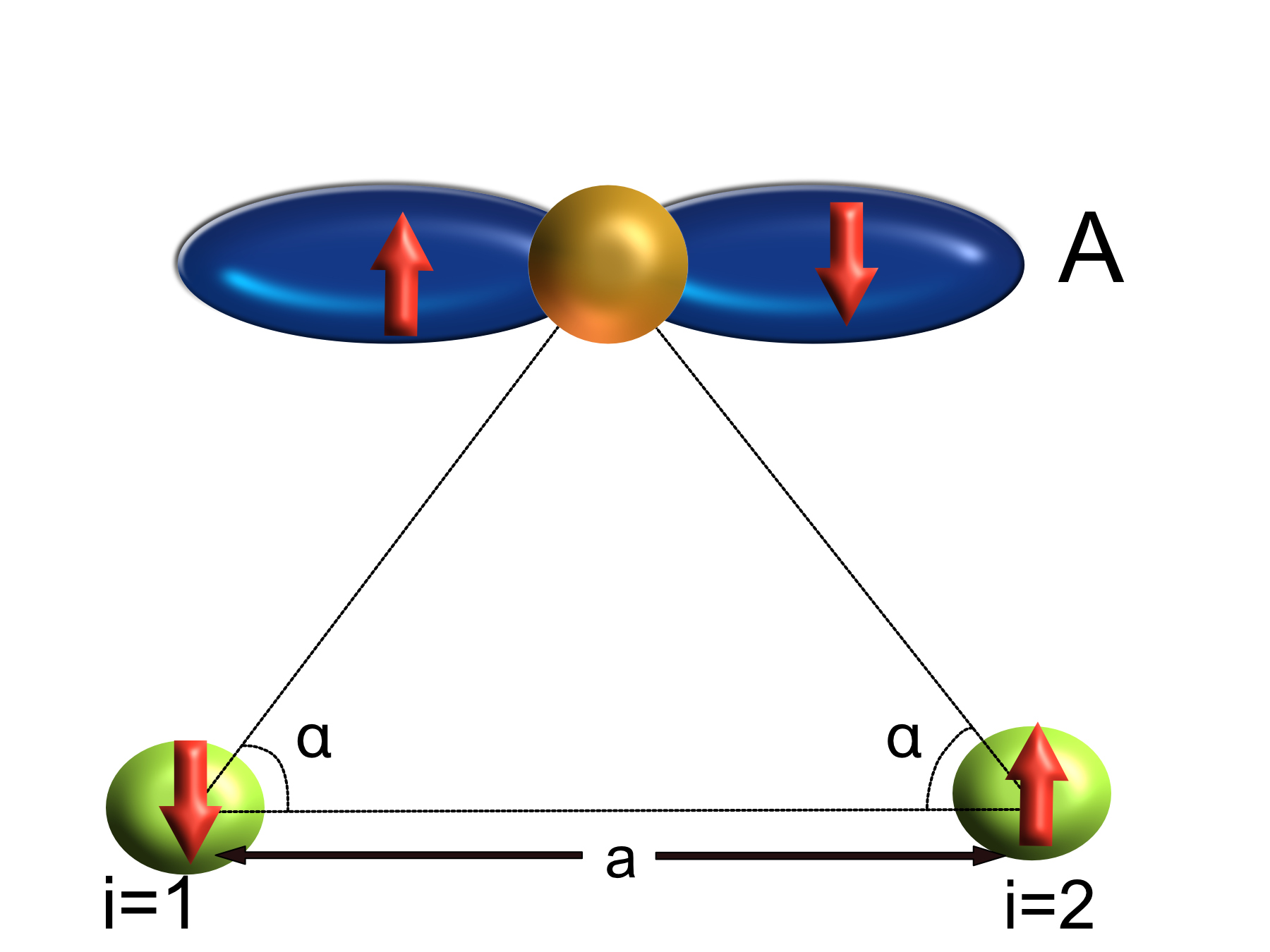}
	\caption{\textbf{AF coupling via one ligand ion.} Spin exchange between two sites(denoted by $i$) with one spin on each  via orbital $A$ of the ligand ion. There is no direct hopping between two metal sites, but the spin can hop between metal site and the orbital $A$ for very small values of bond angle $\alpha$. This superexchange mediated by a non-magnetic ion gives rise to AF interactions between two spins at sites denoted by $i$. }
	\label{oneintermediate}
\end{figure}

\begin{figure}[h]
	\includegraphics[scale=0.48]{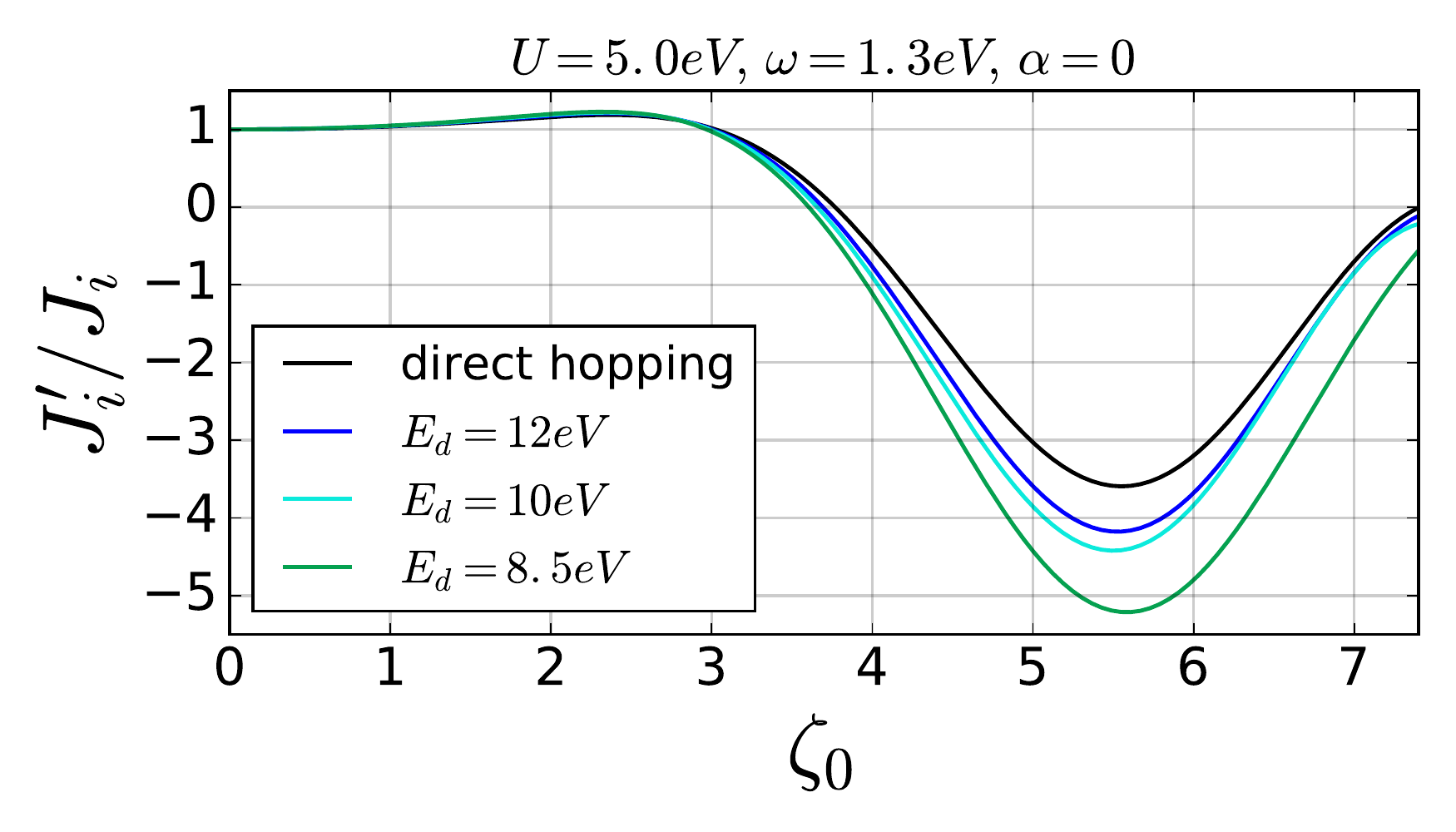}
	\caption{\textbf{Effect of charge transfer gap.} Changes in AFM coupling as a function of drive parameter  for different values of charge transfer gap $E_d$, when the ligand ion lies at the line joining the two TM ions. The qualitative behavior of renormalized coupling is independent of $E_d$, but the quantitative predictions start to differ significantly as $E_d$ decreases. }
	\label{Edchange}
\end{figure}

Consider a simple two-site model with one spin on each site, and with a non-magnetic (ligand) ion between the two metal sites, which mediates the spin exchange between two spins located at metal sites as shown in Fig.~\ref{oneintermediate}. This model can be described the following Hamiltonian:
\begin{equation}
H=-\sum_{i=1,2,\sigma}tc\dg_{i\sigma}c_{A\sigma}+\text{h.c }+E_{A}(\hat{n}_A-2)+U\sum_{i=1,2}\hat{n}_{i\up}\hat{n}_{i\dn},
\end{equation}
where $i$ denotes the metals sites, $A$ denotes the orbitals of non magnetic ions involved in the exchange process and its electronic energy $E_{A}$ is negative, and the on-site interaction on metal sites is $U$. For the undriven case, the magnetic coupling strength is given by:
\begin{equation}
\text{{J}=}4t^{4}\left(\frac{1}{(E_{d})^{2}U}+\frac{1}{(E_{d})^{3}}\right),
\end{equation}
where $E_{d}=|E_{A}|+$U is the energy of those virtual states where one electron has been transferred from the ligand orbital A to the metal ion~\cite{mattis2006theory}.
Now, we apply a uniform AC electric field which adds an extra term $H_{p}$ to above Hamiltonian where,
\begin{equation}
H_{p}=-\sum_{i}e\textbf{E}(t)\cdot\textbf{r}_{i}\hat{n}_{i}+e\textbf{{E}}(t)\cdot\textbf{r}_{A}\hat{n}_{A}
\end{equation}
and $\textbf{E}(t)=E_{0}(\cos\omega t\,\hat{x}+\sin\omega t\,\hat{y})$. Using fourth order perturbation theory, we show that in the non-resonant case, the new coupling strength is modified as follows:

\begin{equation}
\begin{split}
&J_{\text{ex}}	=	4t^{4}\sum_{m_1,n_1,n}\frac{1}{2(E_{d}+n_{1}\omega)(E_{d}+m_{1}\omega)(E_{d}+\frac{n}{2}\omega)}
\\&(\cos(2(n_1-m_1)\alpha)\jj_{m_1}(\zeta_{1})\jj_{n_1}(\zeta_{1})\jj_{n_1-n}(\zeta_{1})\jj_{m_1-n}(\zeta_{1})+\\&
\cos(2(n-m_{1}-n_{1})\alpha)\jj_{n_1}(\zeta_{1})\jj_{m_1}(\zeta_{2})\jj_{n_1-n}(\zeta_{1})\jj_{m_1-n}(\zeta_{2}))\\&
+\frac{1}{(E_{d}+n_{1}\omega)(E_{d}+m_{1}\omega)(U+n\omega)}\\&\left(\cos(2(n_1-m_1)\alpha)\jj_{m_1}(\zeta_{1})\jj_{n_1}(\zeta_{1})\jj_{n_1-n}(\zeta_{2})\jj_{m_1-n}(\zeta_{2})\right)
\end{split}
\end{equation}
where $ \zeta_1=-\zeta_2=\frac{\zeta_{0}}{2\cos\alpha}$, and $\zeta_0=\frac{eE_0a}{\omega}$. The resulting magnetic coupling strength is shown in Fig.~\ref{Edchange} for different values of charge transfer gap $E_d$. We notice that the observed changes do not differ significantly from the direct hopping case if $E_d\gg U$ but the changes for $E_d\approx U$ can be much different.

\begin{figure}[h]

	\includegraphics[scale=0.5]{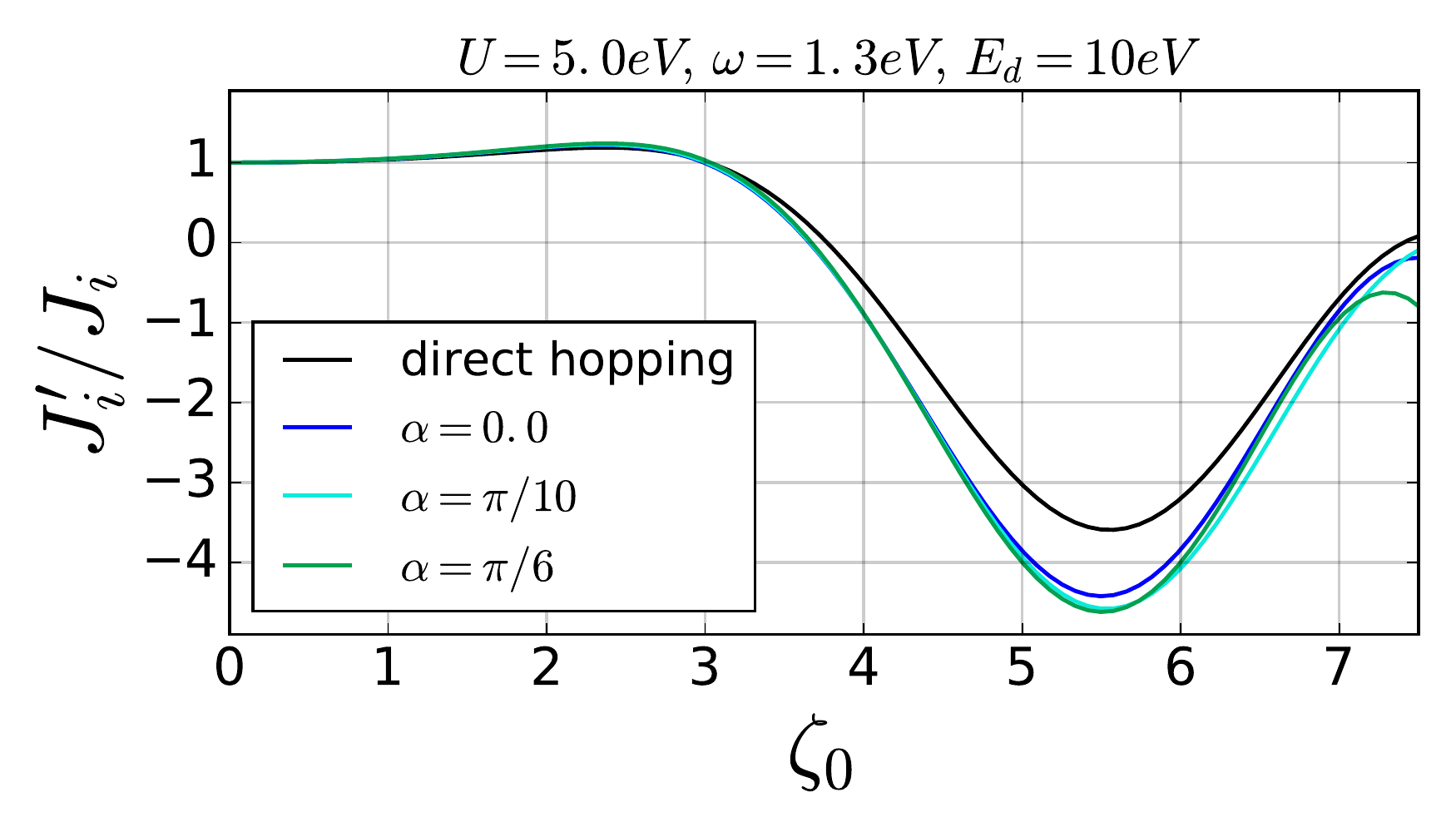}
	\caption{\textbf{Effect of bond angle.} Change in AF coupling as a function of drive parameter for different values of bond angle $\alpha$ in Fig.~(\ref{oneintermediate}). This change follows the same trend as the direct hopping case, and approaches the direct hopping limit when $E_d\gg U,\,\alpha\rightarrow 0$.}
	\label{bondangle}
\end{figure}

In addition to the charge transfer gap, $J_{ex}$ also depends on the bond angle $\alpha$ in Fig.~\ref{oneintermediate}, which is usually close to zero for AF coupling in most cases. In most of the transition metal compounds, only $p$ type orbitals of the ligand ion are involved in this superexchange mechanism, and thus both metal sites should have finite overlap with the same orbital which is possible only if bond angle is very small. We study the effect of periodic drive on magnetic coupling for different bond angles. As shown in Fig.~\ref{bondangle}, the changes in magnetic coupling strength follows the same trend as the direct hopping case approaches the same value when $E_d$ increases and $\alpha\ll1$ .

\subsection*{Case 2 : Effect of a periodic drive on FM coupling mediated by a ligand ion}

\begin{figure}[h]
	\centering
\includegraphics[scale=0.5]{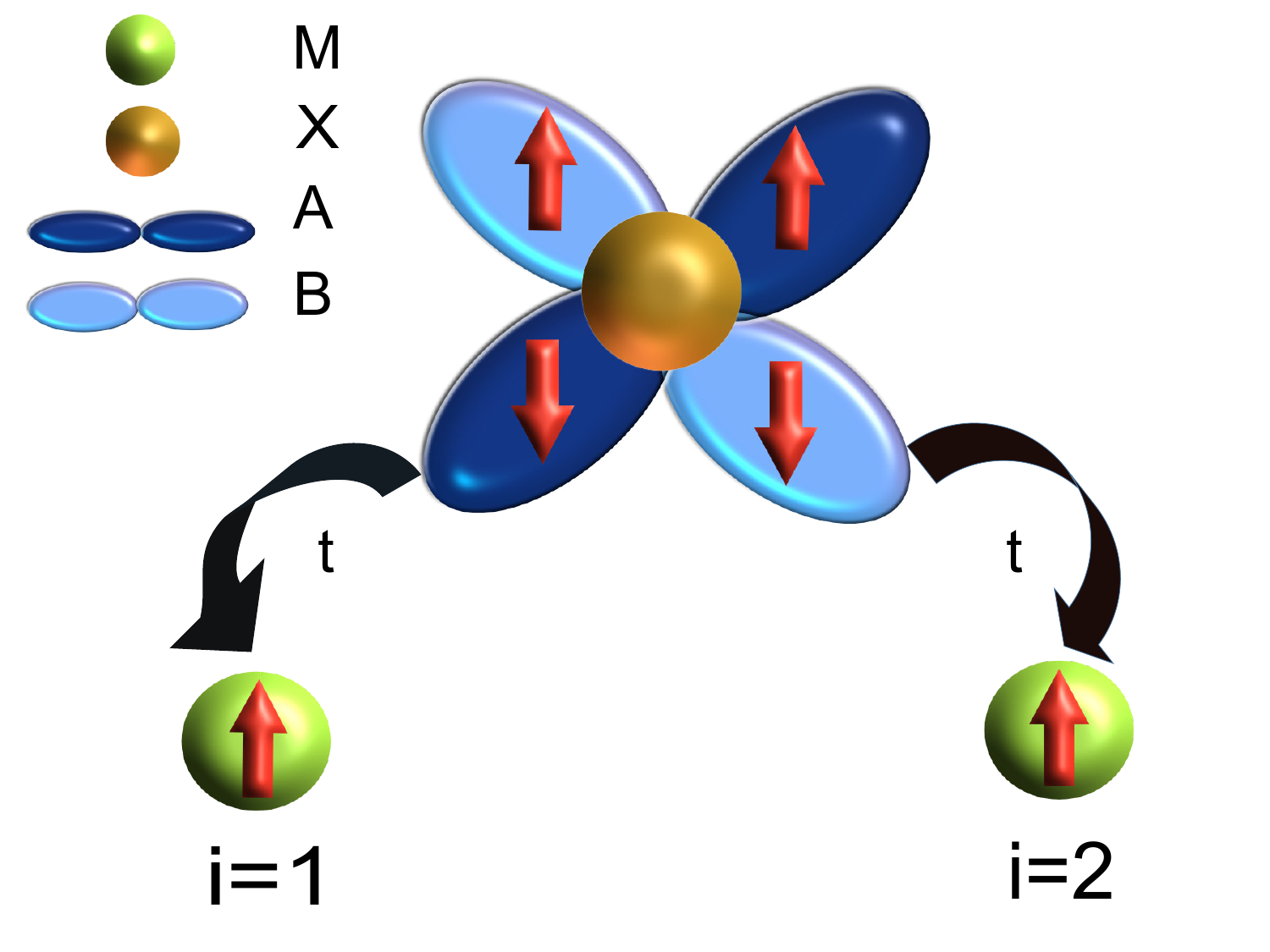}
\caption{\textbf{FM coupling via two orbitals of the ligand ion.} FM interactions between spins mediated by a non-magnetic ion with two orbitals shown in different shades of blue. The electrons from each orbital of the ligand ion can hop to only one metal ion site so the intermediate state with same spin in two orbitals of the ligand ion is preferred due to Hund's coupling. This favors the parallel arrangement of the two spins at metal sites.}
\label{FM}
\end{figure}

In some cases, the ligand ion can also mediate FM interactions. When the spin exchange between two metal sites is not allowed but the two spins can still hop to two different ligand orbitals of the same ligand ion, then due to Hund's coupling, two spins align in the same direction. Even in this situation, the magnetic coupling strength depends on the hopping parameter, and thus can be tuned by a periodic drive to a certain extent. Consider a toy model shown in Fig.~\ref{FM} with two TM ions (M) at sites $i=1$ and $i=2$, and a ligand ion (X) with two degenerate orbitals namely A and B described by Hamiltonian:
\begin{equation}
\begin{split}
H=&U\sum_{i=1,2}\hat{n}_{i\up}\hat{n}_{i\dn}+J_{H}\sum_{\substack{\alpha=A,B,\\ \alpha\ne\alpha'}}c_{\alpha\up} \dg c_{\alpha'\dn}\dg c_{\alpha\dn}c_{\alpha'\up}+\\&\sum_{\substack{\alpha=\{A,B\},\\ \sg=\{\uparrow,\downarrow\}}}E_{A}\hat{n}_{\alpha\sg}-\sum_{\substack{i=\{1,2\},\\ \alpha=\{A,B\}}}t_{i\alpha}(c\dg_{i\sg}c_{\alpha\sg}+c_{\alpha\sg}\dg c_{i\sg})
\end{split}
\end{equation}
where $J_H$ is Hund's coupling and the hopping parameters, $t_{1A}=t_{2B}=t,$ and $t_{2A}=t_{1B}=0$ as the spins at $i=1$, and at $i=2$  hop to orbitals A and B respectively.


\begin{figure}[h]
	\centering
	\includegraphics[scale=0.48]{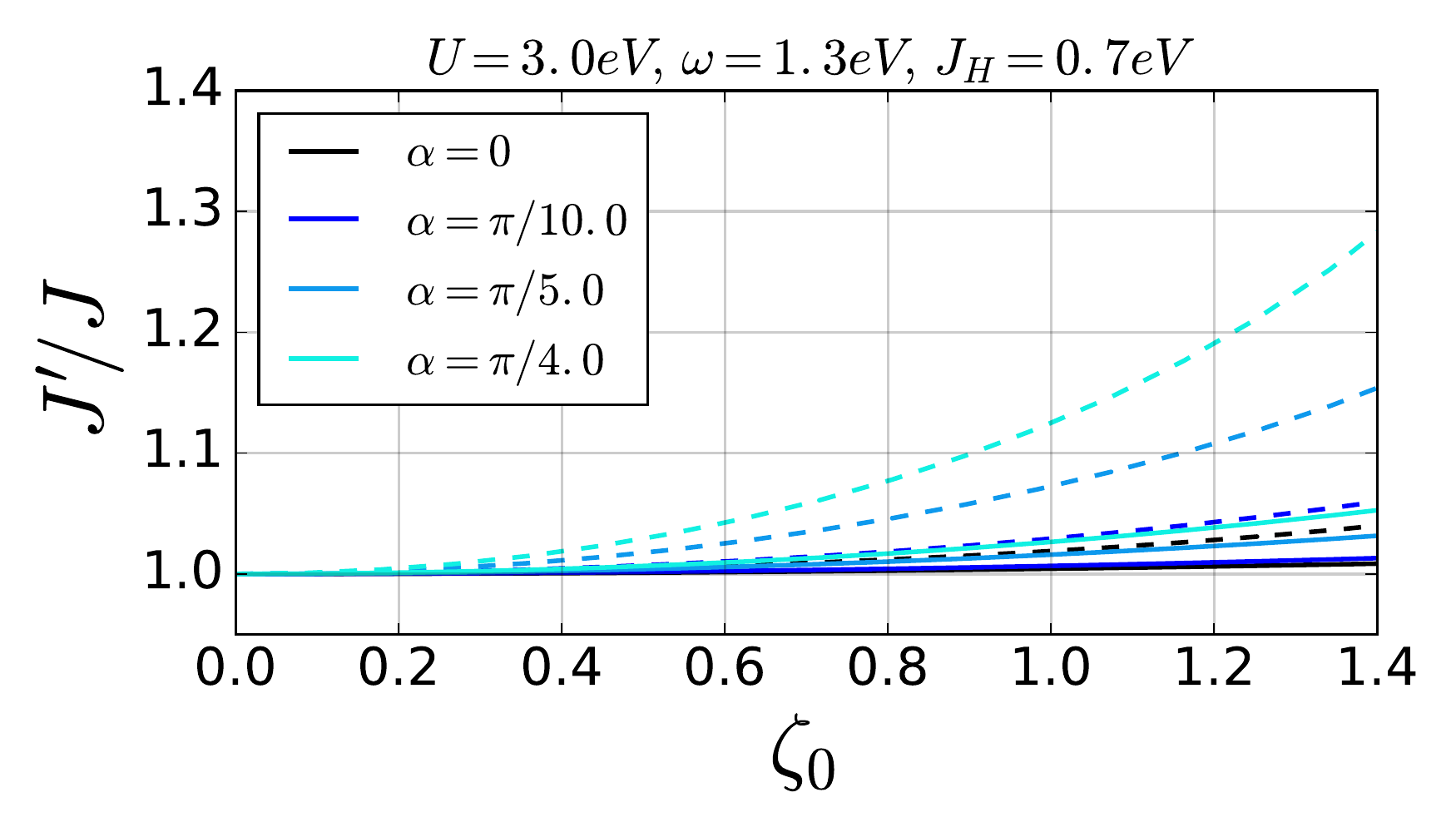}

	\caption{\textbf{Effect of bond angle.}  Change in FM interactions as a function of drive parameter $\zeta_0$ for charge transfer gap $E_d=5eV$(dashed lines) and $E_d=10eV$(solid lines) for different values of angle $\alpha$, where $\alpha$ is the angle between the line joining the two TM ions and the projection of M-X bond on the plane containing these TM ions. In this case, changes depend on angle~$\alpha$ and charge-transfer gap $E_d$. }
	\label{FM change}
\end{figure}
 In the presence of a circularly polarized light, the magnetic coupling strength is given by:
 \begin{equation}
 \begin{split}
 J=t^{4}&\sum_{m_{1},n_{1},n}\frac{4J_{H}}{(E_{d}+m_{1}\omega)(E_{d}+n_{1}\omega)((2E_{d}+n\omega)^{2}-J_{H}^{2})}\\&(\cos(2(m_{1}-n_{1})\alpha)\jj_{n_1}(\zeta_{1a})\jj_{m_1}(\zeta_{1a})\jj_{n-n_1}(\zeta_{2b})\\&\jj_{n-m_1}(\zeta_{2b})+\cos(2(n-m_{1}-n_{1})\alpha)\jj_{n_1}(\zeta_{1a})\\&\jj_{m_1}(\zeta_{2b})\jj_{n-n_1}(\zeta_{2b})\jj_{n-m_1}(\zeta_{1a}))
 \end{split}
 \end{equation}
 where $E_{d}=U-E_{A} ,\, \zeta_{1a}=-\zeta_{2b}=\frac{\zeta_{0}}{2\cos\alpha},$ with $\zeta_{0}=eE_{0}a/\omega$, $a$ is the separation between two TM ions and $\alpha$ is the angle between the line joining the two TM ions and the projection of M-X bond on the plane containing these TM ions. We are using a circularly polarized drive to introduce a direction-independent modification of exchange interactions. As shown in Fig.~\ref{FM change}, the change in FM interaction is sensitive to bond angle $\alpha$ and start to increase with the bond angle. Also, these changes are more significant when the charge transfer gap and drive frequency are of the same order.


\subsection*{Case 3: Effect of periodic drive on AF coupling mediated by two intermediate ligand ions}
\begin{figure}[h]
	\centering
	\includegraphics[scale=0.5]{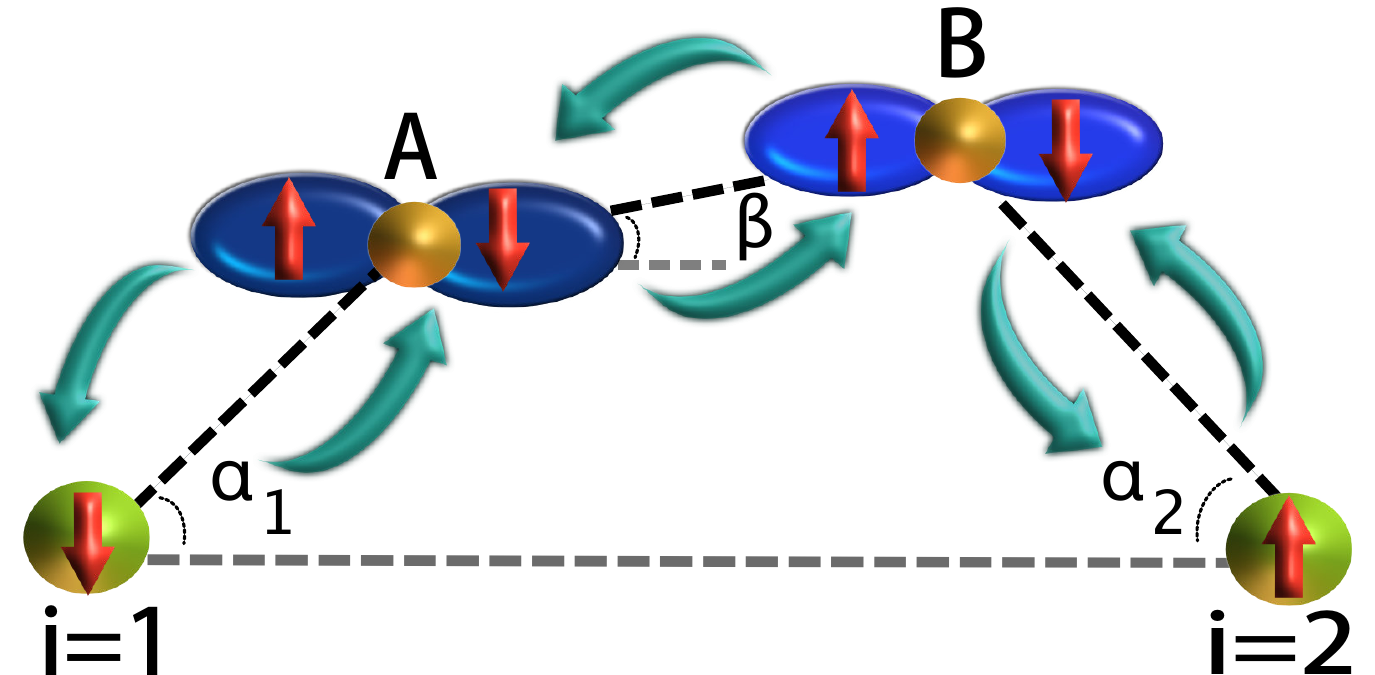}
	\caption{\textbf{AF coupling via two ligand ions.}  AF coupling between two spins mediated by two ligand ions where no direct hopping is allowed between two metal sites. This system is represented by the hamiltonian in Eq.~(\ref{two_ligand}).}
	\label{twoiontoymodel}
\end{figure}

\begin{figure}[h]
	\centering

\includegraphics[scale=0.5]{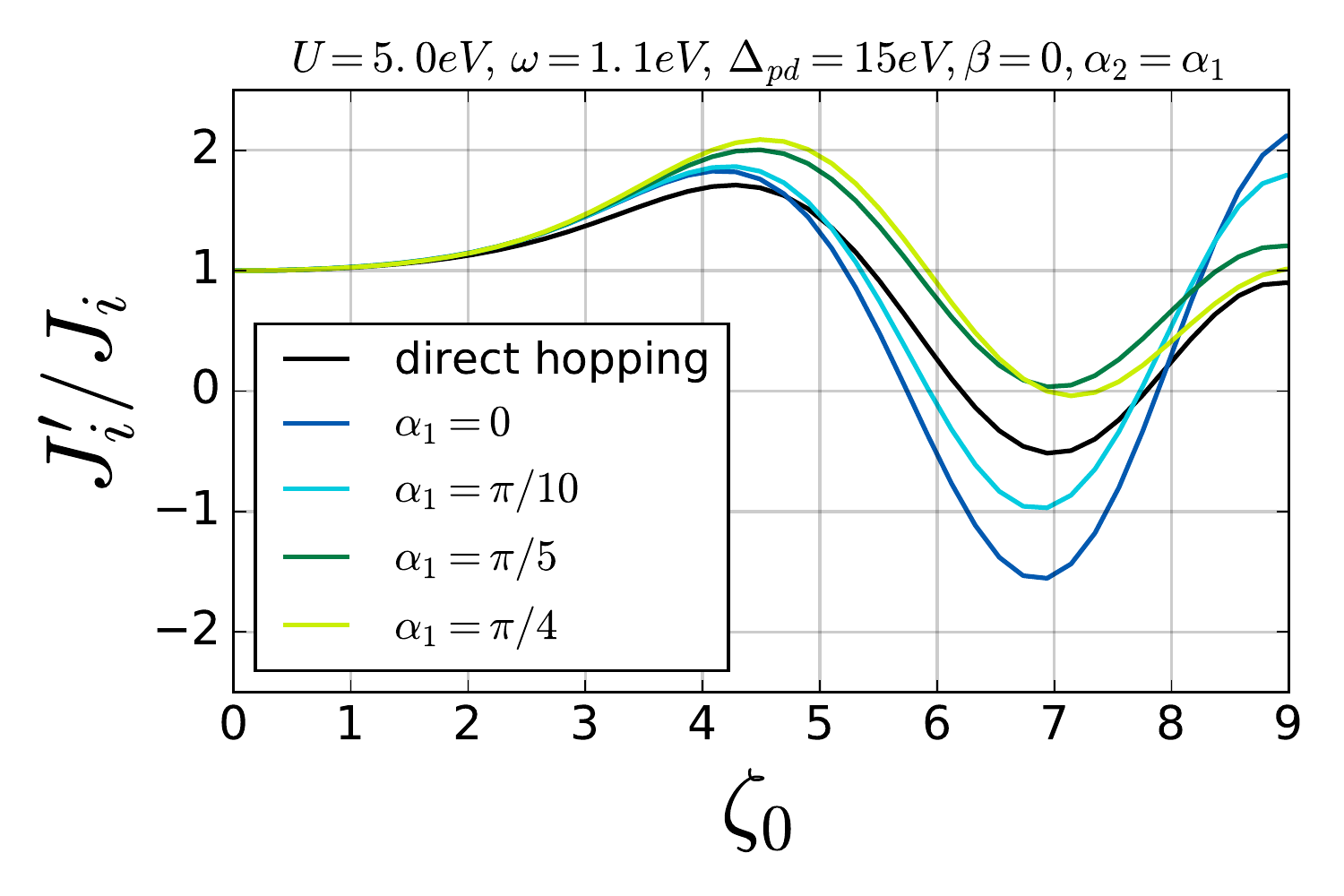}
\caption{\textbf{Effect of bond angles.}  Change in AF coupling strength as a function of drive parameter $\zeta_0=\frac{eEa}{\omega}$, where $a$ is the separation between two magnetic ions and the spin exchange is mediated by two intermediate ions in the presence of a periodic drive. This is one of the most important spin exchange pathway in TMTC monolayer, and the changes in spin exchange interactions are very different from direct hopping case.}
	\label{twoionchange}
\end{figure}

In some materials, especially in TMTC monolayers, second and third nearest neighbor interactions play a very important role in deciding the magnetic ground state. This kind of interactions are allowed only due to the presence of two or more intermediate ions available for spin exchange as shown in Fig.~\ref{exchangepathwayMnPX3}. In such cases, the effect of a periodic drive can be expected to be very different from the direct hopping case. We consider the toy model shown in Fig.~\ref{twoiontoymodel}, where the electron at site $i=1$ can hop to orbital $A$, the electron at other metals site can hop to orbital B of the different ligand ion, and hopping between orbitals $A$ and $B$ is allowed. It can be described by the following hamiltonian:
\begin{equation}
\begin{split}
H=U&\sum_{i}\hat{n}_{i\up}\hat{n}_{i\dn}-\sum_{\substack{i=\{1,2\},\\ \alpha=\{A,B\},\\ \sigma=\{\uparrow,\downarrow\}}}t_{i\alpha}c\dg_{i\sigma}c_{\alpha\sigma}+\text{h.c }+E_{A}(\hat{n}_A-2)\\&+E_{B}(\hat{n}_B-2)-t_{AB}(c\dg_{A\sg}c_{B\sg}+c\dg_{B\sg}c_{A\sg}),
\label{two_ligand}
\end{split}
\end{equation}
where, $E_A$ and $E_B$ are the electronic energies of the orbitals $A$ and $B$ of the ligand ions, $t_{i\alpha}$ denotes the hopping parameter between orbital $\alpha$
 of the ligand ion and metal site $i$, and $t_{1A}=t_{2B}=t$, while $t_{1B}=t_{2A}=0$. Since in most cases, all ligand ions are similar so here we assume $E_A=E_B$. In the presence of a circularly polarized EM field, we calculate the changes in the effective spin-exchange interactions and observe that bond angles play a very important role as shown  in Fig.~\ref{twoionchange}. 

Although Floquet engineering of the spin exchange interactions looks very promising, and changes in the effective spin exchange interactions have been demonstrated in some cold atom experiments~\cite{PhysRevLett.107.210405,gorg2018enhancement}, in real materials, we need very large E fields (of the order of $1V/\AA$) to make any significant changes, and we need to take into account the microscopic details of the spin exchange processes. In most cases, there are more than one competing mechanisms, and quantitatively precise predictions can be made only if the relative contributions from different exchange pathways are known a priori. Indeed, we find that the presence of intermediary ions can not be neglected, and must be taken into account, especially for materials like TMTCs where even the third nearest neighbor interactions play a crucial role in determining the magnetic ground state of the system. This section provided us an estimate of how depending on the exchange mechanism, changes in the magnetic coupling strength can be very different from the direct hopping case.
\begin{figure}
	\centering 
	\includegraphics[scale=0.48]{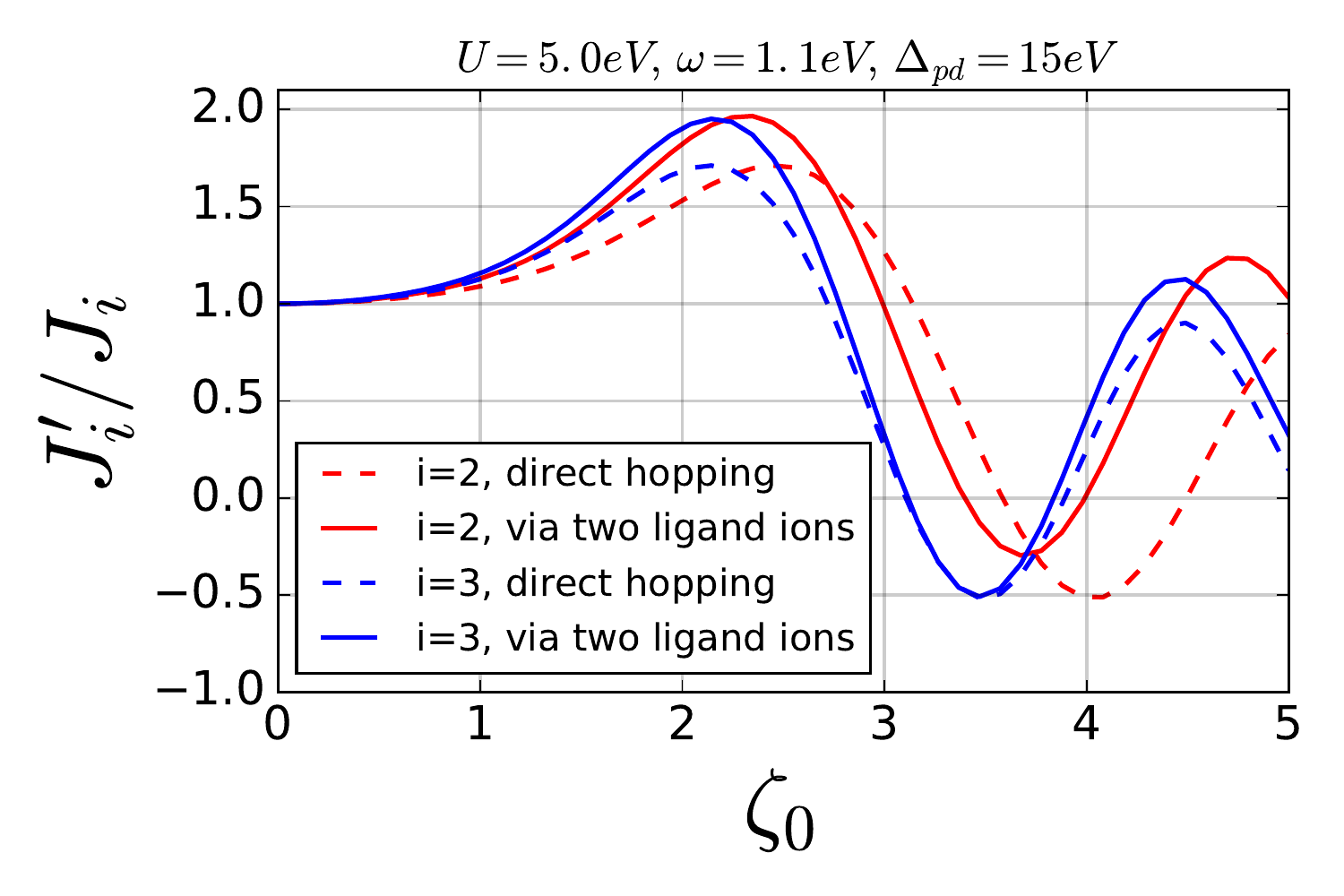}
	\includegraphics[scale=0.48]{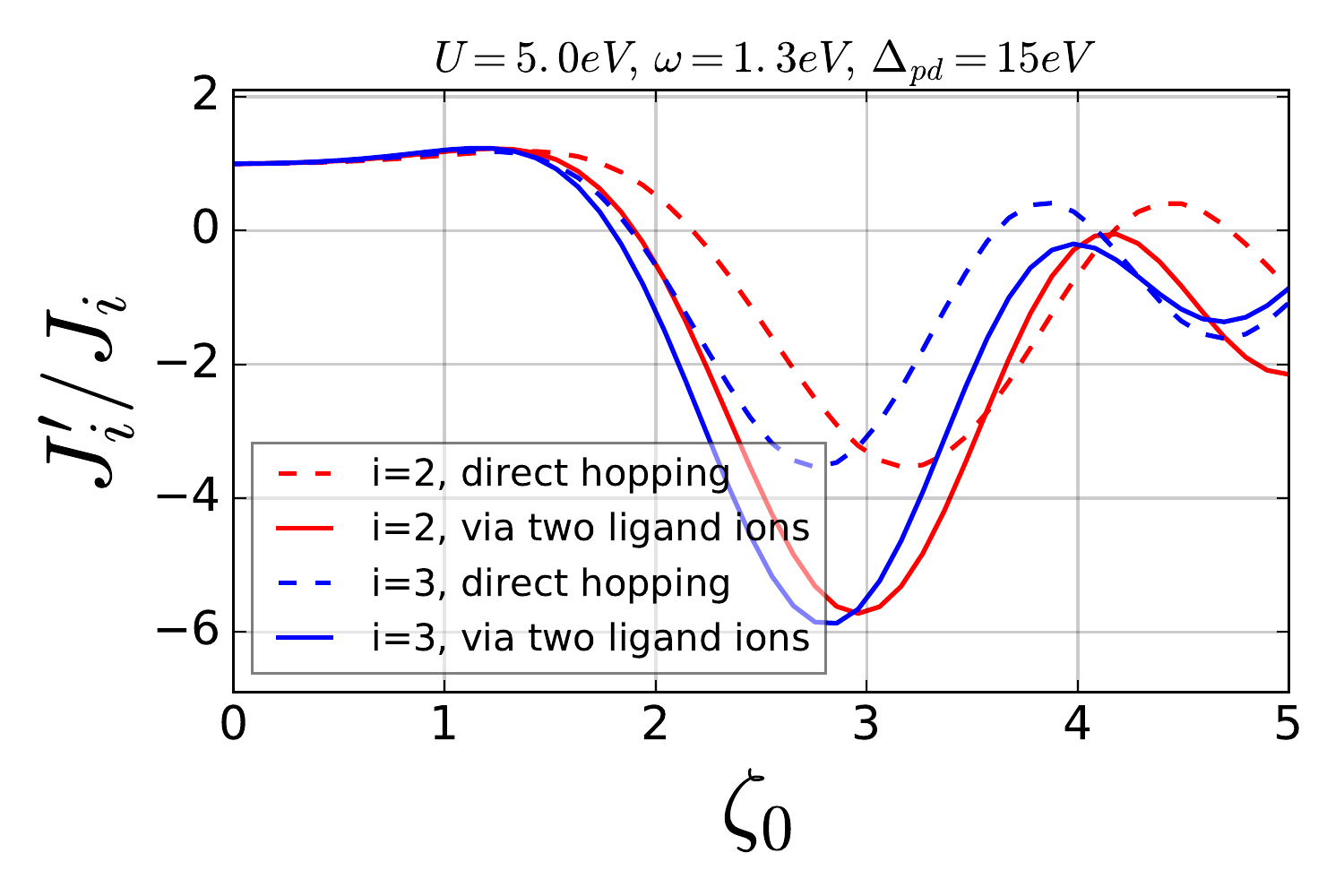}
	\caption{\textbf{Effect of ligands on modified  magnetic coupling in MnPS$_\textbf{3}$.}  Changes in magnetic coupling strength for the second nearest-neighbor ($i=2$) and third nearest neighbor ($i=3$)  as a function of drive parameter $\zeta_0=\frac{eEa}{\omega}$, where $a$ is the distance between nearest-neighbor Mn ions (Here, $a\approx3\AA$ and $\omega\approx1eV$, so electric field $E\approx\frac{\zeta_0}{3}eV/\AA$). We compare the results from a direct hopping model (Fig.~\ref{renormcoupling}) with a more realistic model with ligand ions. These changes were calculated mainly for $\text{MnPS}_3$ in Neel state for the bond parameters taken from Ref.~\cite{chittari_PhysRevB.94.184428}. For the second nearest neighbor case, spin exchange occurs via two different pathways: one involves the two X atoms attached to the same P atom while the other one occurs via two X atoms attached to different P atoms. For the purpose of this calculation, we focused on the first case.}
	\label{MnPS3_Comparison}
\end{figure}

\subsection*{Implications for TMTC monolayers}

The effect of ligand ions on the magnetic coupling renormalization depends on the material properties like bond length, bond angle and charge transfer gap. For TMTC monolayers, the ligand ions mainly affect the second and third nearest neighbor interactions (Fig.~\ref{exchangepathwayMnPX3}). As a result, the analysis shown in Fig.~\ref{renormcoupling} and Fig.~\ref{ratiochange}, where we assumed that all spin exchange processes were occurring as a result of direct hopping would be affected. In Fig.~\ref{MnPS3_Comparison}, we plot the changes in magnetic coupling strength for $\text{MnPS}_3$ by taking into account the different exchange mechanism and compare it to the direct-hopping case. We notice that, depending on the frequency, in certain cases even the qualitative behavior can be drastically different, for, e.g, when the drive frequency is $\omega=1.1eV$, the second and third nearest-neighbor interactions, $J_2$ and $J_3$, change sign at some values of the drive parameter for direct-hopping, but they remain positive when the ligand ions are taken into account. This deviation from the results of direct-hopping model is significant as the ground state of the effective Floquet hamiltonian has a very different magnetic order (Fig.~\ref{ratiochange}) for the above two situations.

\section{Conclusions}

We studied the consequences of a periodic drive on the ligand mediated spin-exchange interactions in a model inspired by TMTCs . Although our calculations are not material specific, they capture the essential features of periodically driven magnetic materials where exchange interactions are mediated by non-magnetic ions. We showed that the modifications due to periodic drive depend on the exchange pathways, and in certain cases these changes can significantly differ from the changes predicted by the direct hopping models. This brings us a step closer to the experimental implementation of Floquet engineering in such materials.

We made several assumptions in order to understand the effects of periodic drive. Particularly, we restricted our analysis to a two-site toy model. Most of our calculations rely on the validity of perturbation methods in Floquet space, and hence we focused on off-resonant cases only. Furthermore, we restricted our analysis to a single orbital on each TM ion except for the FM case. As a result of these simplifications, these findings are applicable to only those cases where degeneracy between different $d$ orbitals is lifted or the exchange process involves only one spin on each magnetic-ion site. 

Also, our discussion was limited to the magnetic properties of Floquet hamiltonians only. In practice, most of the observables also depend on the method used for switching on the drive. This analysis is valid only if the drive is turned on adiabatically. Additionally, the drive must be kept on for a long time to let the system adjust to the new effective hamiltonian. In these spin systems, this time scale is roughly of the order of $1/J\approx5ps$. As we observed in Fig.~\ref{ratiochange} , for certain values of the drive parameter, the effective hamiltonian can have a very different magnetic ground state which can be measured directly from the changes in experimental quantities like reflectivity. While in other cases, the magnetic ground state might not change, but only the strength of magnetic coupling strength is modified. This kind of changes should be reflected in the magnon spectrum or transition temperatures. In certain TMTCs like  $\text{CrSiTe}_\text{3}$, where spin-lattice coupling is very strong~\cite{casto2015strong}, the modified exchange interactions can also affect the phonon frequency shifts which can be studied experimentally.

\textit{Acknowledgements}. We acknowledge support from the IQIM, an NSF Physics Frontiers Center funded through grant PHY-1733907. We are grateful for support from ARO MURI W911NF-16-1-0361 ``Quantum Materials by Design with Electromagnetic Excitation" sponsored by the U.S. Army. GR is also grateful for support from the Simons Foundation and the Packard Foundation. AR is grateful for support from Zuckerman STEM leadership program.

\newpage
\onecolumngrid
\appendix
\section{Derivation for ligand-mediated AF coupling}
\begin{figure}
	\centering 
	
	\includegraphics[scale=0.4]{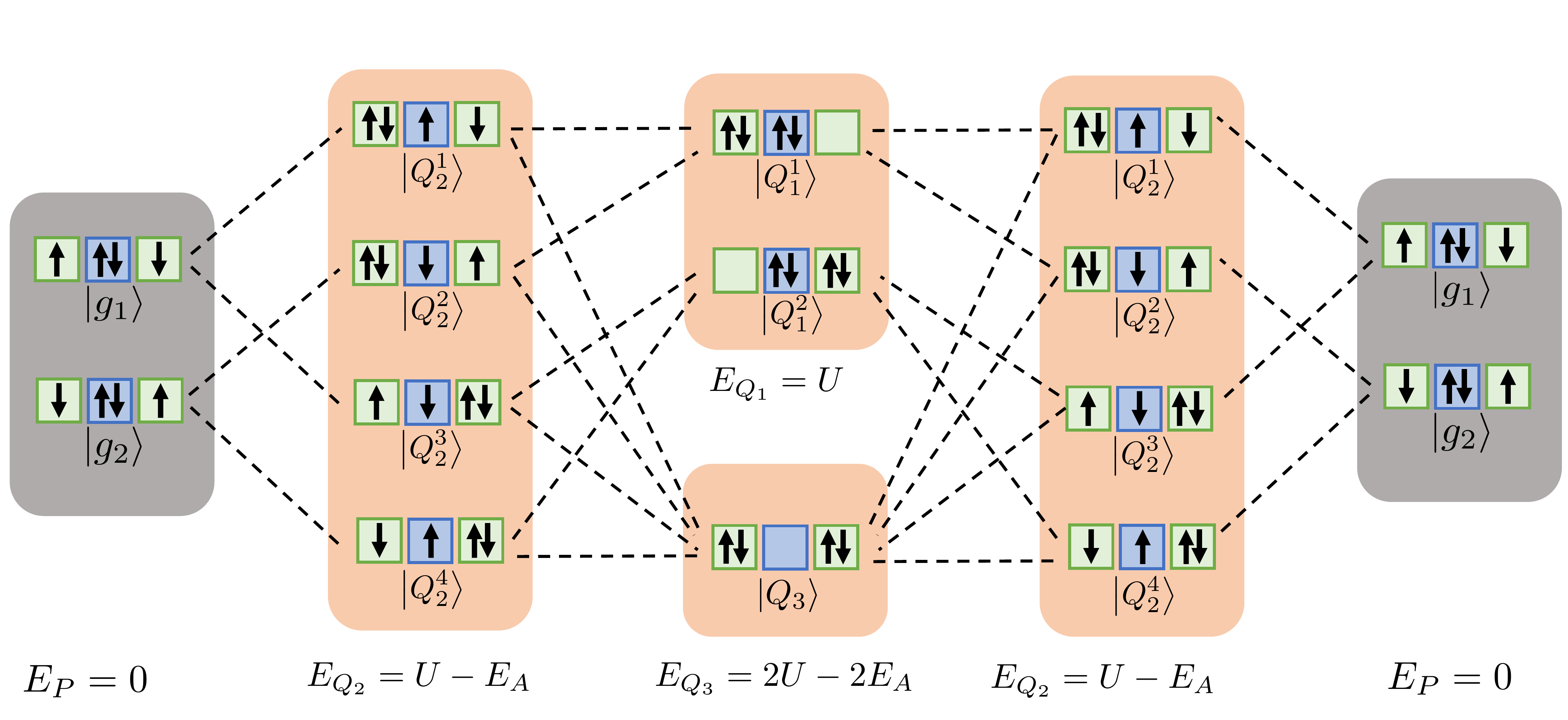}
	\caption{This figure shows the different energy sectors of unperturbed hamiltonian $H_0$ of Eq.~\ref{Happendix}. It also shows all the fourth order perturbation terms which connect the low-energy subspace (shown on left) to the same sub-space (shown on right) after different virtual hoppings to high-energy sectors. All these virtual processes are captured by the effective hamiltonian in $H_\text{eff}^4$ in Eq.~\ref{heff4}. The low energy sector $P$ is shown in gray and  high-energy sectors $Q_1$, $Q_2$, $Q_3$ are shown in orange color. Each state in a given sector is represented by four spins placed in three boxes where green boxes indicate the magnetic ions site and the blue box indicates the ligand-ion site. All the hopping processes are shown by dashed lines. These hopping processes are responsible for the exchange interactions calculated in Eq.~\ref{exchange1}.}
	\label{virtualprocessfig}
\end{figure}

For a two site-model considered in the main text, we have
\begin{equation}
H=E_{A}(\hat{n}_A-2)+U\sum_{i=1,2}\hat{n}_{i\up}\hat{n}_{i\dn}-\sum_{i=1,2,\sigma}t_Ac\dg_{i\sigma}c_{A\sigma}+\text{h.c }=H_0+H_\text{hop},
\label{Happendix}
\end{equation}
where $\hat{n}_A=c\dg_{A\up}c_{A\up}+c\dg_{A\dn}c_{A\dn}$, $i$ denotes the magnetic metal-ion sites, $A$ is the orbital of the non-magnetic ion involved in the exchange process and its electronic energy $E_{A}$ is negative, and the on-site interaction on metal sites is $U$. First, we'll calculate the exchange interactions in the static model where we have one spin on each magnetic ion and the ligand-orbital is completely filled. We can extract the AF coupling by focussing on $S_z=0$ sector only, i.e by finding the energy difference between triplet and singlet configurations. We treat the hopping part as a perturbation. Within $S_z=0$ subspace, $H_0$ has nine eigenstates which can be divided into following four sectors:
\begin{enumerate}
	\item Magnetic-ion single-occupation sector $P$ (two states)

\begin{equation}
\left|g_1\right>=c\dg_{1\uparrow}c\dg_{2\downarrow}c\dg_{A\uparrow}c\dg_{A\downarrow}\left|0\right>,\;\left|g_2\right>=c\dg_{1\downarrow}c\dg_{2\uparrow}c\dg_{A\uparrow}c\dg_{A\downarrow}\left|0\right>
\end{equation}
which has energy $E_P=0$. This also happens to be the low-energy supspace of $H_0$ for $E_A<0$ which is the case here.
	\item Magnetic-ion double-occupation sector $Q_1$ (two states)
	\begin{equation}
	\left|Q_1^1\right>=c\dg_{1\uparrow}c\dg_{1\downarrow}c\dg_{A\uparrow}c\dg_{A\downarrow}\left|0\right>,\;\left|Q_1^2\right>=c\dg_{2\uparrow}c\dg_{2\downarrow}c\dg_{A\uparrow}c\dg_{A\downarrow}\left|0\right>
\end{equation}
with energy $E_{Q_1}=U$.
\item  Ligand-ion single-occupation sector $Q_2$ (four states)
\begin{equation}
\begin{split}
\left|Q_2^1\right>=c\dg_{1\uparrow}c\dg_{1\downarrow}c\dg_{A\uparrow}c\dg_{2\downarrow}\left|0\right>,\,\left|Q_2^2\right>=c\dg_{1\uparrow}c\dg_{1\downarrow}c\dg_{A\downarrow}c\dg_{2\uparrow}\left|0\right>\\
\left|Q_2^3\right>=c\dg_{2\uparrow}c\dg_{2\downarrow}c\dg_{A\downarrow}c\dg_{1\uparrow}\left|0\right>,\,\left|Q_2^4\right>=c\dg_{2\uparrow}c\dg_{2\downarrow}c\dg_{A\uparrow}c\dg_{1\downarrow}\left|0\right>
\end{split}
\end{equation}
which has energy $E_{Q_2}=E_d=U-E_A$.
\item Ligand-ion zero-occupation sector $Q_3$ (one state)
\begin{equation}
\left|Q_3\right>=c\dg_{1\uparrow}c\dg_{1\downarrow}c\dg_{2\uparrow}c\dg_{2\downarrow}\left|0\right>
\end{equation}
which has energy $E_{Q_3}=2E_d=2U-2E_A$.

\end{enumerate}
In order to find the exchange interactions, we can write-down an effective hamiltonian for the ground state sector $P$ by using Schrieffer-Wolf transformations~\cite{bravyi2011schrieffer}. In the model described above, the first contribution to exchange interactions comes from fourth-order corrections in low-energy effective hamiltonian. 
This term arises because of virtual process which connects state $\left|g_1\right>$ to $\left|g_2\right>$ or vice-versa as shown in Fig.~\ref{virtualprocessfig}. The contribution of fourth-order terms can be expressed as:
\begin{equation}
H^4_\text{eff}\approx\sum_{i,j,k=1,2,3}\frac{\hat{P}H_\text{hop}\hat{Q}_iH_\text{hop}\hat{Q}_jH_\text{hop}\hat{Q}_kH_\text{hop}\hat{P}}{(E_P-E_{Q_i})(E_P-E_{Q_j})(E_P-E_{Q_k})}
\label{heff4}
\end{equation}
where $\hat{P}$ and $\hat{Q}_i$ denotes the projection operator on low-energy sector $P$ and high-energy sector $Q_i$.  We notice that 
\begin{equation}
\hat{P}H_\text{hop}\hat{Q}_1=\hat{P}H_\text{hop}\hat{Q}_3=0
\end{equation} which reduces the above sum to
\begin{equation}
H^{(4)}_\text{eff}\approx
-\sum_{j=1,3}\frac{\hat{P}H_\text{hop}\hat{Q}_2H_\text{hop}\hat{Q}_jH_\text{hop}\hat{Q}_2H_\text{hop}\hat{P}}{E_{d}^2E_{Q_j}},
\end{equation}
and it takes the following form
\begin{equation}
H^{(4)}_\text{eff}\approx-\frac{2t_A^4}{E_d^2}\left(\frac{1}{U}+\frac{1}{E_d}\right)\sum_{i,j=1,2}(-1)^{i-j}\ket{g_i}\bra{g_j}
\label{exchange1}
\end{equation}
for states $\ket{g_1},\ket{g_2}$ of low-energy manifold (single-occupation sector $P$) of hamiltonian $H$ in Eq.~\ref{Happendix}. A similar analysis  for $S_z=\pm1$ sector shows that $H^{(4)}_\text{eff}=0$ indicating that the energy of this sector is equal to that of $S_z=0$ triplet state $\frac{1}{\sqrt{2}}\left(\ket{g_1}+\ket{g_2}\right)$ as expected. This gives us a coupling strength:
\begin{equation}
J_\text{ex}=-\frac{4t_A^4}{E_d^2}\left(\frac{1}{U}+\frac{1}{E_d}\right).
\end{equation}
Now, this analysis can be extended to include an  off-resonant drive as well.
For a driven system, after Peierls substitution, $H_\text{hop}$ is replaced by:
\begin{equation}
H_\text{hop}(t)=-\sum_{i=1,2,\sigma}t_Ae^{i\left[\frac{\mathbf{E}(t)\cdot\mathbf{r}_{iA}}{\omega}\right]}c\dg_{i\sigma}c_{A\sigma}+\text{h.c }
\end{equation}
where $\mathbf{r}_{iA}=\mathbf{r}_i-\mathbf{r}_A$ and $\omega$ is the frequency of EM field. For a circularly polarized light, 
\begin{equation}
\begin{split}
\mathbf{E}(t)\cdot\mathbf{r}_{iA}=Er_{iA}\cos(\alpha_{iA})\cos\omega t+ Er_{iA}\sin(\alpha_{iA})\sin\omega t=Er_{iA}\cos\left(\omega t-\alpha_{iA}\right)
\end{split}
\end{equation}
 where $\alpha_{iA}$ is the angle between $\mathbf{r}_{iA}$ and $\hat{x}$ ( we have chosen  $\mathbf{r}_{12}$ as $x$ direction). Now, using Jacob-Anger identity we get
\begin{equation}
H_\text{hop}(t)=-\sum_{i=1,2,\sigma}t_A\jj_{n}(\zeta_{iA})e^{-in\alpha_{iA}}e^{in\omega t}c\dg_{i\sigma}c_{A\sigma}+\text{h.c }
\end{equation}
where $\jj_{n}$ denotes $n^{th}$ order Bessel function and $\zeta_{iA}=Er_{iA}\text{Sign}(x_{iA})$. Now, we can use Floquet theory to express this hamiltonian in a time-independent manner by extendind the original basis to include the photon degree of freedom. In this basis, the hooping part takes the following form
\begin{equation}
H_\text{hop}=\sum_{n,m}H_\text{hop}^n\otimes\ket{m+n}\bra{m}+\text{h.c }
\end{equation}
where $H_\text{hop}^n=-\sum_{i=1,2,\sigma}t_A\jj_{n}(\zeta_{iA})e^{-in\alpha_{iA}}\left(c\dg_{i\sigma}c_{A\sigma}+c\dg_{A\sigma}c_{i\sigma}\right)$ denotes the $n$ photon-assisted hopping,
and similarly the $n^{th}$ sector of unperturbed part becomes
\begin{equation}
H_0^n=(H_0+n\omega)\otimes\ket{n}\bra{n}
\end{equation}
For an off-resonant drive, we can employ the same SW transformation technique in the extended Floquet basis which can account for all photo-assisted virtual processes. Here, again we notice that the lowest-order contribution to exchange splitting comes from the fourth-order terms and it is captured by the following term
\begin{equation}
H^{(4)}_\text{eff}\approx
-\sum_{n_1,n,m_1}\sum_{j=1,3}\frac{\hat{P}H_\text{hop}\hat{Q}_{2,m_1}H^{m_1-n}_\text{hop}\hat{Q}_{j,n}H^{n-n_1}_\text{hop}\hat{Q}_{2,n_1}H^{n_1}_\text{hop}\hat{P}}{(E_{d}+m_1\omega)(E_{Q_j}+n\omega)(E_d+n_1\omega)},
\end{equation}
where $\hat{Q}_{i,n}$ denotes the $n$ photon-dressed $\hat{Q}_i$ sector. By repeating the same steps as in the static case, we get
\begin{equation}
\begin{split}
J_{\text{ex}}	=	4t_A^{4}\sum_{m_1,n_1,n}&\frac{1}{2(E_{d}+n_{1}\omega)(E_{d}+m_{1}\omega)(E_{d}+\frac{n}{2}\omega)}
(\cos(2(n_1-m_1)\alpha)\jj_{m_1}(\zeta_{1})\jj_{n_1}(\zeta_{1})\jj_{n_1-n}(\zeta_{1})\jj_{m_1-n}(\zeta_{1})+\\&
\cos(2(n-m_{1}-n_{1})\alpha)\jj_{n_1}(\zeta_{1})\jj_{m_1}(\zeta_{2})\jj_{n_1-n}(\zeta_{1})\jj_{m_1-n}(\zeta_{2}))+\\&
\frac{1}{(E_{d}+n_{1}\omega)(E_{d}+m_{1}\omega)(U+n\omega)}\left(\cos(2(n_1-m_1)\alpha)\jj_{m_1}(\zeta_{1})\jj_{n_1}(\zeta_{1})\jj_{n_1-n}(\zeta_{2})\jj_{m_1-n}(\zeta_{2})\right)
\end{split}
\end{equation}
where $ \zeta_1=-\zeta_2=\frac{\zeta_{0}}{2\cos\alpha}$, and $\zeta_0=\frac{eE_0a}{\omega}$ where $\alpha=\alpha_{1A}=-\alpha_{2A}$ and $a=|\mathbf{r}_{12}|$.
A similar approach has been used to calculate the magnetic coupling strength for other cases based on an exchange mechanism involving two-ligand orbitals.


\bibliographystyle{apsrev4-1}   
\bibliography{magnetismref.bib}

\begin{thebibliography}{70}%
\makeatletter
\providecommand \@ifxundefined [1]{%
 \@ifx{#1\undefined}
}%
\providecommand \@ifnum [1]{%
 \ifnum #1\expandafter \@firstoftwo
 \else \expandafter \@secondoftwo
 \fi
}%
\providecommand \@ifx [1]{%
 \ifx #1\expandafter \@firstoftwo
 \else \expandafter \@secondoftwo
 \fi
}%
\providecommand \natexlab [1]{#1}%
\providecommand \enquote  [1]{``#1''}%
\providecommand \bibnamefont  [1]{#1}%
\providecommand \bibfnamefont [1]{#1}%
\providecommand \citenamefont [1]{#1}%
\providecommand \href@noop [0]{\@secondoftwo}%
\providecommand \href [0]{\begingroup \@sanitize@url \@href}%
\providecommand \@href[1]{\@@startlink{#1}\@@href}%
\providecommand \@@href[1]{\endgroup#1\@@endlink}%
\providecommand \@sanitize@url [0]{\catcode `\\12\catcode `\$12\catcode
  `\&12\catcode `\#12\catcode `\^12\catcode `\_12\catcode `\%12\relax}%
\providecommand \@@startlink[1]{}%
\providecommand \@@endlink[0]{}%
\providecommand \url  [0]{\begingroup\@sanitize@url \@url }%
\providecommand \@url [1]{\endgroup\@href {#1}{\urlprefix }}%
\providecommand \urlprefix  [0]{URL }%
\providecommand \Eprint [0]{\href }%
\providecommand \doibase [0]{http://dx.doi.org/}%
\providecommand \selectlanguage [0]{\@gobble}%
\providecommand \bibinfo  [0]{\@secondoftwo}%
\providecommand \bibfield  [0]{\@secondoftwo}%
\providecommand \translation [1]{[#1]}%
\providecommand \BibitemOpen [0]{}%
\providecommand \bibitemStop [0]{}%
\providecommand \bibitemNoStop [0]{.\EOS\space}%
\providecommand \EOS [0]{\spacefactor3000\relax}%
\providecommand \BibitemShut  [1]{\csname bibitem#1\endcsname}%
\let\auto@bib@innerbib\@empty
\bibitem [{\citenamefont {Oka}\ and\ \citenamefont
  {Aoki}(2009)}]{F_PhysRevB.79.081406}%
  \BibitemOpen
  \bibfield  {author} {\bibinfo {author} {\bibfnamefont {T.}~\bibnamefont
  {Oka}}\ and\ \bibinfo {author} {\bibfnamefont {H.}~\bibnamefont {Aoki}},\
  }\href {\doibase 10.1103/PhysRevB.79.081406} {\bibfield  {journal} {\bibinfo
  {journal} {Phys. Rev. B}\ }\textbf {\bibinfo {volume} {79}},\ \bibinfo
  {pages} {081406} (\bibinfo {year} {2009})}\BibitemShut {NoStop}%
\bibitem [{\citenamefont {Kitagawa}\ \emph {et~al.}(2011)\citenamefont
  {Kitagawa}, \citenamefont {Oka}, \citenamefont {Brataas}, \citenamefont
  {Fu},\ and\ \citenamefont {Demler}}]{F_PhysRevB.84.235108}%
  \BibitemOpen
  \bibfield  {author} {\bibinfo {author} {\bibfnamefont {T.}~\bibnamefont
  {Kitagawa}}, \bibinfo {author} {\bibfnamefont {T.}~\bibnamefont {Oka}},
  \bibinfo {author} {\bibfnamefont {A.}~\bibnamefont {Brataas}}, \bibinfo
  {author} {\bibfnamefont {L.}~\bibnamefont {Fu}}, \ and\ \bibinfo {author}
  {\bibfnamefont {E.}~\bibnamefont {Demler}},\ }\href {\doibase
  10.1103/PhysRevB.84.235108} {\bibfield  {journal} {\bibinfo  {journal} {Phys.
  Rev. B}\ }\textbf {\bibinfo {volume} {84}},\ \bibinfo {pages} {235108}
  (\bibinfo {year} {2011})}\BibitemShut {NoStop}%
\bibitem [{\citenamefont {Inoue}\ and\ \citenamefont
  {Tanaka}(2010)}]{F_PhysRevLett.105.017401}%
  \BibitemOpen
  \bibfield  {author} {\bibinfo {author} {\bibfnamefont {J.-i.}\ \bibnamefont
  {Inoue}}\ and\ \bibinfo {author} {\bibfnamefont {A.}~\bibnamefont {Tanaka}},\
  }\href {\doibase 10.1103/PhysRevLett.105.017401} {\bibfield  {journal}
  {\bibinfo  {journal} {Phys. Rev. Lett.}\ ,\ \bibinfo {pages} {017401}}
  (\bibinfo {year} {2010})}\BibitemShut {NoStop}%
\bibitem [{\citenamefont {Lindner}\ \emph {et~al.}(2011)\citenamefont
  {Lindner}, \citenamefont {Refael},\ and\ \citenamefont
  {Galitski}}]{F_lindner2011floquet}%
  \BibitemOpen
  \bibfield  {author} {\bibinfo {author} {\bibfnamefont {N.~H.}\ \bibnamefont
  {Lindner}}, \bibinfo {author} {\bibfnamefont {G.}~\bibnamefont {Refael}}, \
  and\ \bibinfo {author} {\bibfnamefont {V.}~\bibnamefont {Galitski}},\ }\href
  {https://www.nature.com/articles/nphys1926} {\bibfield  {journal} {\bibinfo
  {journal} {Nat. Phys.}\ }\textbf {\bibinfo {volume} {7}},\ \bibinfo {pages}
  {490} (\bibinfo {year} {2011})}\BibitemShut {NoStop}%
\bibitem [{\citenamefont {Goldman}\ and\ \citenamefont
  {Dalibard}(2014)}]{F_PhysRevX.4.031027}%
  \BibitemOpen
  \bibfield  {author} {\bibinfo {author} {\bibfnamefont {N.}~\bibnamefont
  {Goldman}}\ and\ \bibinfo {author} {\bibfnamefont {J.}~\bibnamefont
  {Dalibard}},\ }\href {\doibase 10.1103/PhysRevX.4.031027} {\bibfield
  {journal} {\bibinfo  {journal} {Phys. Rev. X}\ }\textbf {\bibinfo {volume}
  {4}},\ \bibinfo {pages} {031027} (\bibinfo {year} {2014})}\BibitemShut
  {NoStop}%
\bibitem [{\citenamefont {Titum}\ \emph {et~al.}(2016)\citenamefont {Titum},
  \citenamefont {Berg}, \citenamefont {Rudner}, \citenamefont {Refael},\ and\
  \citenamefont {Lindner}}]{F_PhysRevX.6.021013}%
  \BibitemOpen
  \bibfield  {author} {\bibinfo {author} {\bibfnamefont {P.}~\bibnamefont
  {Titum}}, \bibinfo {author} {\bibfnamefont {E.}~\bibnamefont {Berg}},
  \bibinfo {author} {\bibfnamefont {M.~S.}\ \bibnamefont {Rudner}}, \bibinfo
  {author} {\bibfnamefont {G.}~\bibnamefont {Refael}}, \ and\ \bibinfo {author}
  {\bibfnamefont {N.~H.}\ \bibnamefont {Lindner}},\ }\href {\doibase
  10.1103/PhysRevX.6.021013} {\bibfield  {journal} {\bibinfo  {journal} {Phys.
  Rev. X}\ }\textbf {\bibinfo {volume} {6}},\ \bibinfo {pages} {021013}
  (\bibinfo {year} {2016})}\BibitemShut {NoStop}%
\bibitem [{\citenamefont {Ezawa}(2013)}]{F_PhysRevLett.110.026603}%
  \BibitemOpen
  \bibfield  {author} {\bibinfo {author} {\bibfnamefont {M.}~\bibnamefont
  {Ezawa}},\ }\href {\doibase 10.1103/PhysRevLett.110.026603} {\bibfield
  {journal} {\bibinfo  {journal} {Phys. Rev. Lett.}\ }\textbf {\bibinfo
  {volume} {110}},\ \bibinfo {pages} {026603} (\bibinfo {year}
  {2013})}\BibitemShut {NoStop}%
\bibitem [{\citenamefont {Bairey}\ \emph {et~al.}(2017)\citenamefont {Bairey},
  \citenamefont {Refael},\ and\ \citenamefont
  {Lindner}}]{F2_PhysRevB.96.020201}%
  \BibitemOpen
  \bibfield  {author} {\bibinfo {author} {\bibfnamefont {E.}~\bibnamefont
  {Bairey}}, \bibinfo {author} {\bibfnamefont {G.}~\bibnamefont {Refael}}, \
  and\ \bibinfo {author} {\bibfnamefont {N.~H.}\ \bibnamefont {Lindner}},\
  }\href {\doibase 10.1103/PhysRevB.96.020201} {\bibfield  {journal} {\bibinfo
  {journal} {Phys. Rev. B}\ }\textbf {\bibinfo {volume} {96}},\ \bibinfo
  {pages} {020201} (\bibinfo {year} {2017})}\BibitemShut {NoStop}%
\bibitem [{\citenamefont {Fausti}\ \emph {et~al.}(2011)\citenamefont {Fausti},
  \citenamefont {Tobey}, \citenamefont {Dean}, \citenamefont {Kaiser},
  \citenamefont {Dienst}, \citenamefont {Hoffmann}, \citenamefont {Pyon},
  \citenamefont {Takayama}, \citenamefont {Takagi},\ and\ \citenamefont
  {Cavalleri}}]{F_fausti2011light}%
  \BibitemOpen
  \bibfield  {author} {\bibinfo {author} {\bibfnamefont {D.}~\bibnamefont
  {Fausti}}, \bibinfo {author} {\bibfnamefont {R.}~\bibnamefont {Tobey}},
  \bibinfo {author} {\bibfnamefont {N.}~\bibnamefont {Dean}}, \bibinfo {author}
  {\bibfnamefont {S.}~\bibnamefont {Kaiser}}, \bibinfo {author} {\bibfnamefont
  {A.}~\bibnamefont {Dienst}}, \bibinfo {author} {\bibfnamefont {M.~C.}\
  \bibnamefont {Hoffmann}}, \bibinfo {author} {\bibfnamefont {S.}~\bibnamefont
  {Pyon}}, \bibinfo {author} {\bibfnamefont {T.}~\bibnamefont {Takayama}},
  \bibinfo {author} {\bibfnamefont {H.}~\bibnamefont {Takagi}}, \ and\ \bibinfo
  {author} {\bibfnamefont {A.}~\bibnamefont {Cavalleri}},\ }\href
  {http://science.sciencemag.org/content/sci/331/6014/189.full.pdf?sid=7ab2aa67-2776-4d79-accd-2670edd5f0f4}
  {\bibfield  {journal} {\bibinfo  {journal} {science}\ }\textbf {\bibinfo
  {volume} {331}},\ \bibinfo {pages} {189} (\bibinfo {year}
  {2011})}\BibitemShut {NoStop}%
\bibitem [{\citenamefont {Klinovaja}\ \emph {et~al.}(2016)\citenamefont
  {Klinovaja}, \citenamefont {Stano},\ and\ \citenamefont
  {Loss}}]{F_PhysRevLett.116.176401}%
  \BibitemOpen
  \bibfield  {author} {\bibinfo {author} {\bibfnamefont {J.}~\bibnamefont
  {Klinovaja}}, \bibinfo {author} {\bibfnamefont {P.}~\bibnamefont {Stano}}, \
  and\ \bibinfo {author} {\bibfnamefont {D.}~\bibnamefont {Loss}},\ }\href
  {\doibase 10.1103/PhysRevLett.116.176401} {\bibfield  {journal} {\bibinfo
  {journal} {Phys. Rev. Lett.}\ }\textbf {\bibinfo {volume} {116}},\ \bibinfo
  {pages} {176401} (\bibinfo {year} {2016})}\BibitemShut {NoStop}%
\bibitem [{\citenamefont {Grushin}\ \emph {et~al.}(2014)\citenamefont
  {Grushin}, \citenamefont {G\'omez-Le\'on},\ and\ \citenamefont
  {Neupert}}]{F_PhysRevLett.112.156801}%
  \BibitemOpen
  \bibfield  {author} {\bibinfo {author} {\bibfnamefont {A.~G.}\ \bibnamefont
  {Grushin}}, \bibinfo {author} {\bibfnamefont {A.}~\bibnamefont
  {G\'omez-Le\'on}}, \ and\ \bibinfo {author} {\bibfnamefont {T.}~\bibnamefont
  {Neupert}},\ }\href {\doibase 10.1103/PhysRevLett.112.156801} {\bibfield
  {journal} {\bibinfo  {journal} {Phys. Rev. Lett.}\ }\textbf {\bibinfo
  {volume} {112}},\ \bibinfo {pages} {156801} (\bibinfo {year}
  {2014})}\BibitemShut {NoStop}%
\bibitem [{\citenamefont {Hasan}\ \emph
  {et~al.}(2017{\natexlab{a}})\citenamefont {Hasan}, \citenamefont {Yudin},
  \citenamefont {Iorsh}, \citenamefont {Eriksson},\ and\ \citenamefont
  {Shelykh}}]{F2_PhysRevB.96.205127}%
  \BibitemOpen
  \bibfield  {author} {\bibinfo {author} {\bibfnamefont {M.}~\bibnamefont
  {Hasan}}, \bibinfo {author} {\bibfnamefont {D.}~\bibnamefont {Yudin}},
  \bibinfo {author} {\bibfnamefont {I.}~\bibnamefont {Iorsh}}, \bibinfo
  {author} {\bibfnamefont {O.}~\bibnamefont {Eriksson}}, \ and\ \bibinfo
  {author} {\bibfnamefont {I.}~\bibnamefont {Shelykh}},\ }\href {\doibase
  10.1103/PhysRevB.96.205127} {\bibfield  {journal} {\bibinfo  {journal} {Phys.
  Rev. B}\ }\textbf {\bibinfo {volume} {96}},\ \bibinfo {pages} {205127}
  (\bibinfo {year} {2017}{\natexlab{a}})}\BibitemShut {NoStop}%
\bibitem [{\citenamefont {Hauke}\ \emph {et~al.}(2012)\citenamefont {Hauke},
  \citenamefont {Tieleman}, \citenamefont {Celi}, \citenamefont
  {\"Olschl\"ager}, \citenamefont {Simonet}, \citenamefont {Struck},
  \citenamefont {Weinberg}, \citenamefont {Windpassinger}, \citenamefont
  {Sengstock}, \citenamefont {Lewenstein},\ and\ \citenamefont
  {Eckardt}}]{F2_PhysRevLett.109.145301}%
  \BibitemOpen
  \bibfield  {author} {\bibinfo {author} {\bibfnamefont {P.}~\bibnamefont
  {Hauke}}, \bibinfo {author} {\bibfnamefont {O.}~\bibnamefont {Tieleman}},
  \bibinfo {author} {\bibfnamefont {A.}~\bibnamefont {Celi}}, \bibinfo {author}
  {\bibfnamefont {C.}~\bibnamefont {\"Olschl\"ager}}, \bibinfo {author}
  {\bibfnamefont {J.}~\bibnamefont {Simonet}}, \bibinfo {author} {\bibfnamefont
  {J.}~\bibnamefont {Struck}}, \bibinfo {author} {\bibfnamefont
  {M.}~\bibnamefont {Weinberg}}, \bibinfo {author} {\bibfnamefont
  {P.}~\bibnamefont {Windpassinger}}, \bibinfo {author} {\bibfnamefont
  {K.}~\bibnamefont {Sengstock}}, \bibinfo {author} {\bibfnamefont
  {M.}~\bibnamefont {Lewenstein}}, \ and\ \bibinfo {author} {\bibfnamefont
  {A.}~\bibnamefont {Eckardt}},\ }\href {\doibase
  10.1103/PhysRevLett.109.145301} {\bibfield  {journal} {\bibinfo  {journal}
  {Phys. Rev. Lett.}\ }\textbf {\bibinfo {volume} {109}},\ \bibinfo {pages}
  {145301} (\bibinfo {year} {2012})}\BibitemShut {NoStop}%
\bibitem [{\citenamefont {Harper}\ and\ \citenamefont
  {Roy}(2017)}]{F2_PhysRevLett.118.115301}%
  \BibitemOpen
  \bibfield  {author} {\bibinfo {author} {\bibfnamefont {F.}~\bibnamefont
  {Harper}}\ and\ \bibinfo {author} {\bibfnamefont {R.}~\bibnamefont {Roy}},\
  }\href {\doibase 10.1103/PhysRevLett.118.115301} {\bibfield  {journal}
  {\bibinfo  {journal} {Phys. Rev. Lett.}\ }\textbf {\bibinfo {volume} {118}},\
  \bibinfo {pages} {115301} (\bibinfo {year} {2017})}\BibitemShut {NoStop}%
\bibitem [{\citenamefont {Baum}\ \emph {et~al.}(2018)\citenamefont {Baum},
  \citenamefont {van Nieuwenburg},\ and\ \citenamefont
  {Refael}}]{F2_baum2018dynamical}%
  \BibitemOpen
  \bibfield  {author} {\bibinfo {author} {\bibfnamefont {Y.}~\bibnamefont
  {Baum}}, \bibinfo {author} {\bibfnamefont {E.~P.}\ \bibnamefont {van
  Nieuwenburg}}, \ and\ \bibinfo {author} {\bibfnamefont {G.}~\bibnamefont
  {Refael}},\ }\href {https://arxiv.org/pdf/1802.08262.pdf} {\bibfield
  {journal} {\bibinfo  {journal} {arXiv preprint arXiv:1802.08262}\ } (\bibinfo
  {year} {2018})}\BibitemShut {NoStop}%
\bibitem [{\citenamefont {Goldman}\ \emph {et~al.}(2014)\citenamefont
  {Goldman}, \citenamefont {Juzeli{\=u}nas}, \citenamefont {{\"O}hberg},\ and\
  \citenamefont {Spielman}}]{F2_goldman2014light}%
  \BibitemOpen
  \bibfield  {author} {\bibinfo {author} {\bibfnamefont {N.}~\bibnamefont
  {Goldman}}, \bibinfo {author} {\bibfnamefont {G.}~\bibnamefont
  {Juzeli{\=u}nas}}, \bibinfo {author} {\bibfnamefont {P.}~\bibnamefont
  {{\"O}hberg}}, \ and\ \bibinfo {author} {\bibfnamefont {I.~B.}\ \bibnamefont
  {Spielman}},\ }\href
  {http://iopscience.iop.org/article/10.1088/0034-4885/77/12/126401/meta}
  {\bibfield  {journal} {\bibinfo  {journal} {Reports on Progress in Physics}\
  }\textbf {\bibinfo {volume} {77}},\ \bibinfo {pages} {126401} (\bibinfo
  {year} {2014})}\BibitemShut {NoStop}%
\bibitem [{\citenamefont {Holthaus}(2015)}]{F2_holthaus2015floquet}%
  \BibitemOpen
  \bibfield  {author} {\bibinfo {author} {\bibfnamefont {M.}~\bibnamefont
  {Holthaus}},\ }\href
  {http://iopscience.iop.org/article/10.1088/0953-4075/49/1/013001/meta}
  {\bibfield  {journal} {\bibinfo  {journal} {Journal of Physics B: Atomic,
  Molecular and Optical Physics}\ }\textbf {\bibinfo {volume} {49}},\ \bibinfo
  {pages} {013001} (\bibinfo {year} {2015})}\BibitemShut {NoStop}%
\bibitem [{\citenamefont {Itin}\ and\ \citenamefont
  {Katsnelson}(2015)}]{F2_itin2015effective}%
  \BibitemOpen
  \bibfield  {author} {\bibinfo {author} {\bibfnamefont {A.}~\bibnamefont
  {Itin}}\ and\ \bibinfo {author} {\bibfnamefont {M.}~\bibnamefont
  {Katsnelson}},\ }\href
  {https://journals.aps.org/prl/abstract/10.1103/PhysRevLett.115.075301}
  {\bibfield  {journal} {\bibinfo  {journal} {Phys. Rev. Lett.}\ }\textbf
  {\bibinfo {volume} {115}},\ \bibinfo {pages} {075301} (\bibinfo {year}
  {2015})}\BibitemShut {NoStop}%
\bibitem [{\citenamefont {Mishra}\ \emph {et~al.}(2018)\citenamefont {Mishra},
  \citenamefont {Pallaprolu}, \citenamefont {Guha~Sarkar},\ and\ \citenamefont
  {Bandyopadhyay}}]{FK_PhysRevB.97.085405}%
  \BibitemOpen
  \bibfield  {author} {\bibinfo {author} {\bibfnamefont {T.}~\bibnamefont
  {Mishra}}, \bibinfo {author} {\bibfnamefont {A.}~\bibnamefont {Pallaprolu}},
  \bibinfo {author} {\bibfnamefont {T.}~\bibnamefont {Guha~Sarkar}}, \ and\
  \bibinfo {author} {\bibfnamefont {J.~N.}\ \bibnamefont {Bandyopadhyay}},\
  }\href {\doibase 10.1103/PhysRevB.97.085405} {\bibfield  {journal} {\bibinfo
  {journal} {Phys. Rev. B}\ }\textbf {\bibinfo {volume} {97}},\ \bibinfo
  {pages} {085405} (\bibinfo {year} {2018})}\BibitemShut {NoStop}%
\bibitem [{\citenamefont {Wang}\ \emph {et~al.}(2017)\citenamefont {Wang},
  \citenamefont {Li},\ and\ \citenamefont {Li}}]{FK_PhysRevB.95.104308}%
  \BibitemOpen
  \bibfield  {author} {\bibinfo {author} {\bibfnamefont {L.~C.}\ \bibnamefont
  {Wang}}, \bibinfo {author} {\bibfnamefont {X.~P.}\ \bibnamefont {Li}}, \ and\
  \bibinfo {author} {\bibfnamefont {C.~F.}\ \bibnamefont {Li}},\ }\href
  {\doibase 10.1103/PhysRevB.95.104308} {\bibfield  {journal} {\bibinfo
  {journal} {Phys. Rev. B}\ }\textbf {\bibinfo {volume} {95}},\ \bibinfo
  {pages} {104308} (\bibinfo {year} {2017})}\BibitemShut {NoStop}%
\bibitem [{\citenamefont {Plekhanov}\ \emph {et~al.}(2017)\citenamefont
  {Plekhanov}, \citenamefont {Roux},\ and\ \citenamefont
  {Le~Hur}}]{F_PhysRevB.95.045102}%
  \BibitemOpen
  \bibfield  {author} {\bibinfo {author} {\bibfnamefont {K.}~\bibnamefont
  {Plekhanov}}, \bibinfo {author} {\bibfnamefont {G.}~\bibnamefont {Roux}}, \
  and\ \bibinfo {author} {\bibfnamefont {K.}~\bibnamefont {Le~Hur}},\ }\href
  {\doibase 10.1103/PhysRevB.95.045102} {\bibfield  {journal} {\bibinfo
  {journal} {Phys. Rev. B}\ }\textbf {\bibinfo {volume} {95}},\ \bibinfo
  {pages} {045102} (\bibinfo {year} {2017})}\BibitemShut {NoStop}%
\bibitem [{\citenamefont {Roman-Taboada}\ and\ \citenamefont
  {Naumis}(2017)}]{FE_PhysRevB.96.155435}%
  \BibitemOpen
  \bibfield  {author} {\bibinfo {author} {\bibfnamefont {P.}~\bibnamefont
  {Roman-Taboada}}\ and\ \bibinfo {author} {\bibfnamefont {G.~G.}\ \bibnamefont
  {Naumis}},\ }\href {\doibase 10.1103/PhysRevB.96.155435} {\bibfield
  {journal} {\bibinfo  {journal} {Phys. Rev. B}\ }\textbf {\bibinfo {volume}
  {96}},\ \bibinfo {pages} {155435} (\bibinfo {year} {2017})}\BibitemShut
  {NoStop}%
\bibitem [{\citenamefont {Takasan}\ \emph
  {et~al.}(2017{\natexlab{a}})\citenamefont {Takasan}, \citenamefont {Daido},
  \citenamefont {Kawakami},\ and\ \citenamefont
  {Yanase}}]{FM_PhysRevB.95.134508}%
  \BibitemOpen
  \bibfield  {author} {\bibinfo {author} {\bibfnamefont {K.}~\bibnamefont
  {Takasan}}, \bibinfo {author} {\bibfnamefont {A.}~\bibnamefont {Daido}},
  \bibinfo {author} {\bibfnamefont {N.}~\bibnamefont {Kawakami}}, \ and\
  \bibinfo {author} {\bibfnamefont {Y.}~\bibnamefont {Yanase}},\ }\href
  {\doibase 10.1103/PhysRevB.95.134508} {\bibfield  {journal} {\bibinfo
  {journal} {Phys. Rev. B}\ }\textbf {\bibinfo {volume} {95}},\ \bibinfo
  {pages} {134508} (\bibinfo {year} {2017}{\natexlab{a}})}\BibitemShut
  {NoStop}%
\bibitem [{\citenamefont {Thakurathi}\ \emph {et~al.}(2017)\citenamefont
  {Thakurathi}, \citenamefont {Loss},\ and\ \citenamefont
  {Klinovaja}}]{FM_PhysRevB.95.155407}%
  \BibitemOpen
  \bibfield  {author} {\bibinfo {author} {\bibfnamefont {M.}~\bibnamefont
  {Thakurathi}}, \bibinfo {author} {\bibfnamefont {D.}~\bibnamefont {Loss}}, \
  and\ \bibinfo {author} {\bibfnamefont {J.}~\bibnamefont {Klinovaja}},\ }\href
  {\doibase 10.1103/PhysRevB.95.155407} {\bibfield  {journal} {\bibinfo
  {journal} {Phys. Rev. B}\ }\textbf {\bibinfo {volume} {95}},\ \bibinfo
  {pages} {155407} (\bibinfo {year} {2017})}\BibitemShut {NoStop}%
\bibitem [{\citenamefont {Benito}\ \emph {et~al.}(2014)\citenamefont {Benito},
  \citenamefont {G\'omez-Le\'on}, \citenamefont {Bastidas}, \citenamefont
  {Brandes},\ and\ \citenamefont {Platero}}]{FM_PhysRevB.90.205127}%
  \BibitemOpen
  \bibfield  {author} {\bibinfo {author} {\bibfnamefont {M.}~\bibnamefont
  {Benito}}, \bibinfo {author} {\bibfnamefont {A.}~\bibnamefont
  {G\'omez-Le\'on}}, \bibinfo {author} {\bibfnamefont {V.~M.}\ \bibnamefont
  {Bastidas}}, \bibinfo {author} {\bibfnamefont {T.}~\bibnamefont {Brandes}}, \
  and\ \bibinfo {author} {\bibfnamefont {G.}~\bibnamefont {Platero}},\ }\href
  {\doibase 10.1103/PhysRevB.90.205127} {\bibfield  {journal} {\bibinfo
  {journal} {Phys. Rev. B}\ }\textbf {\bibinfo {volume} {90}},\ \bibinfo
  {pages} {205127} (\bibinfo {year} {2014})}\BibitemShut {NoStop}%
\bibitem [{\citenamefont {Li}\ \emph {et~al.}(2017)\citenamefont {Li},
  \citenamefont {Lam},\ and\ \citenamefont {You}}]{FM_PhysRevB.96.155438}%
  \BibitemOpen
  \bibfield  {author} {\bibinfo {author} {\bibfnamefont {Z.-Z.}\ \bibnamefont
  {Li}}, \bibinfo {author} {\bibfnamefont {C.-H.}\ \bibnamefont {Lam}}, \ and\
  \bibinfo {author} {\bibfnamefont {J.~Q.}\ \bibnamefont {You}},\ }\href
  {\doibase 10.1103/PhysRevB.96.155438} {\bibfield  {journal} {\bibinfo
  {journal} {Phys. Rev. B}\ }\textbf {\bibinfo {volume} {96}},\ \bibinfo
  {pages} {155438} (\bibinfo {year} {2017})}\BibitemShut {NoStop}%
\bibitem [{\citenamefont {Yang}\ \emph {et~al.}(2018)\citenamefont {Yang},
  \citenamefont {Huang},\ and\ \citenamefont {Wang}}]{FM_yang2018floquet}%
  \BibitemOpen
  \bibfield  {author} {\bibinfo {author} {\bibfnamefont {X.}~\bibnamefont
  {Yang}}, \bibinfo {author} {\bibfnamefont {B.}~\bibnamefont {Huang}}, \ and\
  \bibinfo {author} {\bibfnamefont {Z.}~\bibnamefont {Wang}},\ }\href
  {https://www.nature.com/articles/s41598-018-20604-w} {\bibfield  {journal}
  {\bibinfo  {journal} {Scientific reports}\ }\textbf {\bibinfo {volume} {8}},\
  \bibinfo {pages} {2243} (\bibinfo {year} {2018})}\BibitemShut {NoStop}%
\bibitem [{\citenamefont {Saha}\ \emph {et~al.}(2017)\citenamefont {Saha},
  \citenamefont {Sivarajan},\ and\ \citenamefont
  {Sen}}]{FM_PhysRevB.95.174306}%
  \BibitemOpen
  \bibfield  {author} {\bibinfo {author} {\bibfnamefont {S.}~\bibnamefont
  {Saha}}, \bibinfo {author} {\bibfnamefont {S.~N.}\ \bibnamefont {Sivarajan}},
  \ and\ \bibinfo {author} {\bibfnamefont {D.}~\bibnamefont {Sen}},\ }\href
  {\doibase 10.1103/PhysRevB.95.174306} {\bibfield  {journal} {\bibinfo
  {journal} {Phys. Rev. B}\ }\textbf {\bibinfo {volume} {95}},\ \bibinfo
  {pages} {174306} (\bibinfo {year} {2017})}\BibitemShut {NoStop}%
\bibitem [{\citenamefont {Jotzu}\ \emph {et~al.}(2014)\citenamefont {Jotzu},
  \citenamefont {Messer}, \citenamefont {Desbuquois}, \citenamefont {Lebrat},
  \citenamefont {Uehlinger}, \citenamefont {Greif},\ and\ \citenamefont
  {Esslinger}}]{FSL_jotzu2014experimental}%
  \BibitemOpen
  \bibfield  {author} {\bibinfo {author} {\bibfnamefont {G.}~\bibnamefont
  {Jotzu}}, \bibinfo {author} {\bibfnamefont {M.}~\bibnamefont {Messer}},
  \bibinfo {author} {\bibfnamefont {R.}~\bibnamefont {Desbuquois}}, \bibinfo
  {author} {\bibfnamefont {M.}~\bibnamefont {Lebrat}}, \bibinfo {author}
  {\bibfnamefont {T.}~\bibnamefont {Uehlinger}}, \bibinfo {author}
  {\bibfnamefont {D.}~\bibnamefont {Greif}}, \ and\ \bibinfo {author}
  {\bibfnamefont {T.}~\bibnamefont {Esslinger}},\ }\href
  {https://www.nature.com/articles/nature13915} {\bibfield  {journal} {\bibinfo
   {journal} {Nature}\ }\textbf {\bibinfo {volume} {515}},\ \bibinfo {pages}
  {237} (\bibinfo {year} {2014})}\BibitemShut {NoStop}%
\bibitem [{\citenamefont {Ding}\ \emph {et~al.}(2018)\citenamefont {Ding},
  \citenamefont {Lim}, \citenamefont {Su},\ and\ \citenamefont
  {Weng}}]{FQ_PhysRevB.97.035123}%
  \BibitemOpen
  \bibfield  {author} {\bibinfo {author} {\bibfnamefont {K.-H.}\ \bibnamefont
  {Ding}}, \bibinfo {author} {\bibfnamefont {L.-K.}\ \bibnamefont {Lim}},
  \bibinfo {author} {\bibfnamefont {G.}~\bibnamefont {Su}}, \ and\ \bibinfo
  {author} {\bibfnamefont {Z.-Y.}\ \bibnamefont {Weng}},\ }\href {\doibase
  10.1103/PhysRevB.97.035123} {\bibfield  {journal} {\bibinfo  {journal} {Phys.
  Rev. B}\ }\textbf {\bibinfo {volume} {97}},\ \bibinfo {pages} {035123}
  (\bibinfo {year} {2018})}\BibitemShut {NoStop}%
\bibitem [{\citenamefont {Owerre}(2018)}]{MagFl_owerre2018photoinduced}%
  \BibitemOpen
  \bibfield  {author} {\bibinfo {author} {\bibfnamefont {S.}~\bibnamefont
  {Owerre}},\ }\href {https://www.nature.com/articles/s41598-018-22779-8}
  {\bibfield  {journal} {\bibinfo  {journal} {Scientific reports}\ }\textbf
  {\bibinfo {volume} {8}},\ \bibinfo {pages} {4431} (\bibinfo {year}
  {2018})}\BibitemShut {NoStop}%
\bibitem [{\citenamefont {Hasan}\ \emph
  {et~al.}(2017{\natexlab{b}})\citenamefont {Hasan}, \citenamefont {Yudin},
  \citenamefont {Iorsh}, \citenamefont {Eriksson},\ and\ \citenamefont
  {Shelykh}}]{F0_PhysRevB.96.205127}%
  \BibitemOpen
  \bibfield  {author} {\bibinfo {author} {\bibfnamefont {M.}~\bibnamefont
  {Hasan}}, \bibinfo {author} {\bibfnamefont {D.}~\bibnamefont {Yudin}},
  \bibinfo {author} {\bibfnamefont {I.}~\bibnamefont {Iorsh}}, \bibinfo
  {author} {\bibfnamefont {O.}~\bibnamefont {Eriksson}}, \ and\ \bibinfo
  {author} {\bibfnamefont {I.}~\bibnamefont {Shelykh}},\ }\href {\doibase
  10.1103/PhysRevB.96.205127} {\bibfield  {journal} {\bibinfo  {journal} {Phys.
  Rev. B}\ }\textbf {\bibinfo {volume} {96}},\ \bibinfo {pages} {205127}
  (\bibinfo {year} {2017}{\natexlab{b}})}\BibitemShut {NoStop}%
\bibitem [{\citenamefont {Agarwala}\ and\ \citenamefont
  {Sen}(2017)}]{F0_PhysRevB.96.104309}%
  \BibitemOpen
  \bibfield  {author} {\bibinfo {author} {\bibfnamefont {A.}~\bibnamefont
  {Agarwala}}\ and\ \bibinfo {author} {\bibfnamefont {D.}~\bibnamefont {Sen}},\
  }\href {\doibase 10.1103/PhysRevB.96.104309} {\bibfield  {journal} {\bibinfo
  {journal} {Phys. Rev. B}\ }\textbf {\bibinfo {volume} {96}},\ \bibinfo
  {pages} {104309} (\bibinfo {year} {2017})}\BibitemShut {NoStop}%
\bibitem [{\citenamefont {Pervishko}\ \emph {et~al.}(2018)\citenamefont
  {Pervishko}, \citenamefont {Yudin},\ and\ \citenamefont
  {Shelykh}}]{F0_PhysRevB.97.075420}%
  \BibitemOpen
  \bibfield  {author} {\bibinfo {author} {\bibfnamefont {A.~A.}\ \bibnamefont
  {Pervishko}}, \bibinfo {author} {\bibfnamefont {D.}~\bibnamefont {Yudin}}, \
  and\ \bibinfo {author} {\bibfnamefont {I.~A.}\ \bibnamefont {Shelykh}},\
  }\href {\doibase 10.1103/PhysRevB.97.075420} {\bibfield  {journal} {\bibinfo
  {journal} {Phys. Rev. B}\ }\textbf {\bibinfo {volume} {97}},\ \bibinfo
  {pages} {075420} (\bibinfo {year} {2018})}\BibitemShut {NoStop}%
\bibitem [{\citenamefont {Thuberg}\ \emph {et~al.}(2017)\citenamefont
  {Thuberg}, \citenamefont {Mu\~noz}, \citenamefont {Eggert},\ and\
  \citenamefont {Reyes}}]{F0_PhysRevLett.119.267701}%
  \BibitemOpen
  \bibfield  {author} {\bibinfo {author} {\bibfnamefont {D.}~\bibnamefont
  {Thuberg}}, \bibinfo {author} {\bibfnamefont {E.}~\bibnamefont {Mu\~noz}},
  \bibinfo {author} {\bibfnamefont {S.}~\bibnamefont {Eggert}}, \ and\ \bibinfo
  {author} {\bibfnamefont {S.~A.}\ \bibnamefont {Reyes}},\ }\href {\doibase
  10.1103/PhysRevLett.119.267701} {\bibfield  {journal} {\bibinfo  {journal}
  {Phys. Rev. Lett.}\ }\textbf {\bibinfo {volume} {119}},\ \bibinfo {pages}
  {267701} (\bibinfo {year} {2017})}\BibitemShut {NoStop}%
\bibitem [{\citenamefont {Inoue}(2018)}]{F0_inoue2018floquet}%
  \BibitemOpen
  \bibfield  {author} {\bibinfo {author} {\bibfnamefont {J.-i.}\ \bibnamefont
  {Inoue}},\ }\href {https://journals.jps.jp/doi/10.7566/JPSJ.87.034711}
  {\bibfield  {journal} {\bibinfo  {journal} {Journal of the Physical Society
  of Japan}\ }\textbf {\bibinfo {volume} {87}},\ \bibinfo {pages} {034711}
  (\bibinfo {year} {2018})}\BibitemShut {NoStop}%
\bibitem [{\citenamefont {Takasan}\ \emph
  {et~al.}(2017{\natexlab{b}})\citenamefont {Takasan}, \citenamefont
  {Nakagawa},\ and\ \citenamefont {Kawakami}}]{FKondo_PhysRevB.96.115120}%
  \BibitemOpen
  \bibfield  {author} {\bibinfo {author} {\bibfnamefont {K.}~\bibnamefont
  {Takasan}}, \bibinfo {author} {\bibfnamefont {M.}~\bibnamefont {Nakagawa}}, \
  and\ \bibinfo {author} {\bibfnamefont {N.}~\bibnamefont {Kawakami}},\ }\href
  {\doibase 10.1103/PhysRevB.96.115120} {\bibfield  {journal} {\bibinfo
  {journal} {Phys. Rev. B}\ }\textbf {\bibinfo {volume} {96}},\ \bibinfo
  {pages} {115120} (\bibinfo {year} {2017}{\natexlab{b}})}\BibitemShut
  {NoStop}%
\bibitem [{\citenamefont {Zhang}\ \emph {et~al.}(2016)\citenamefont {Zhang},
  \citenamefont {Ong},\ and\ \citenamefont
  {Nagaosa}}]{Fweyl_PhysRevB.94.235137}%
  \BibitemOpen
  \bibfield  {author} {\bibinfo {author} {\bibfnamefont {X.-X.}\ \bibnamefont
  {Zhang}}, \bibinfo {author} {\bibfnamefont {T.~T.}\ \bibnamefont {Ong}}, \
  and\ \bibinfo {author} {\bibfnamefont {N.}~\bibnamefont {Nagaosa}},\ }\href
  {\doibase 10.1103/PhysRevB.94.235137} {\bibfield  {journal} {\bibinfo
  {journal} {Phys. Rev. B}\ }\textbf {\bibinfo {volume} {94}},\ \bibinfo
  {pages} {235137} (\bibinfo {year} {2016})}\BibitemShut {NoStop}%
\bibitem [{\citenamefont {Chan}\ \emph {et~al.}(2016)\citenamefont {Chan},
  \citenamefont {Oh}, \citenamefont {Han},\ and\ \citenamefont
  {Lee}}]{FWeyl_PhysRevB.94.121106}%
  \BibitemOpen
  \bibfield  {author} {\bibinfo {author} {\bibfnamefont {C.-K.}\ \bibnamefont
  {Chan}}, \bibinfo {author} {\bibfnamefont {Y.-T.}\ \bibnamefont {Oh}},
  \bibinfo {author} {\bibfnamefont {J.~H.}\ \bibnamefont {Han}}, \ and\
  \bibinfo {author} {\bibfnamefont {P.~A.}\ \bibnamefont {Lee}},\ }\href
  {\doibase 10.1103/PhysRevB.94.121106} {\bibfield  {journal} {\bibinfo
  {journal} {Phys. Rev. B}\ }\textbf {\bibinfo {volume} {94}},\ \bibinfo
  {pages} {121106} (\bibinfo {year} {2016})}\BibitemShut {NoStop}%
\bibitem [{\citenamefont {Fu}\ \emph {et~al.}(2017)\citenamefont {Fu},
  \citenamefont {Duan}, \citenamefont {Wang},\ and\ \citenamefont
  {Chen}}]{FWeyl_fu2017phase}%
  \BibitemOpen
  \bibfield  {author} {\bibinfo {author} {\bibfnamefont {P.-H.}\ \bibnamefont
  {Fu}}, \bibinfo {author} {\bibfnamefont {H.-J.}\ \bibnamefont {Duan}},
  \bibinfo {author} {\bibfnamefont {R.-Q.}\ \bibnamefont {Wang}}, \ and\
  \bibinfo {author} {\bibfnamefont {H.}~\bibnamefont {Chen}},\ }\href
  {https://www.sciencedirect.com/science/article/pii/S0375960117308149?via%3Dihub}
  {\bibfield  {journal} {\bibinfo  {journal} {Physics Letters A}\ }\textbf
  {\bibinfo {volume} {381}},\ \bibinfo {pages} {3499} (\bibinfo {year}
  {2017})}\BibitemShut {NoStop}%
\bibitem [{\citenamefont {Niu}\ and\ \citenamefont
  {Wu}(2018)}]{Fweyl_niu2018tunable}%
  \BibitemOpen
  \bibfield  {author} {\bibinfo {author} {\bibfnamefont {Z.~P.}\ \bibnamefont
  {Niu}}\ and\ \bibinfo {author} {\bibfnamefont {X.}~\bibnamefont {Wu}},\
  }\href {https://www.sciencedirect.com/science/article/pii/S0375960118300070}
  {\bibfield  {journal} {\bibinfo  {journal} {Physics Letters A}\ } (\bibinfo
  {year} {2018})}\BibitemShut {NoStop}%
\bibitem [{\citenamefont
  {Ezawa}(2017{\natexlab{a}})}]{Fweyl_PhysRevB.96.041205}%
  \BibitemOpen
  \bibfield  {author} {\bibinfo {author} {\bibfnamefont {M.}~\bibnamefont
  {Ezawa}},\ }\href {\doibase 10.1103/PhysRevB.96.041205} {\bibfield  {journal}
  {\bibinfo  {journal} {Phys. Rev. B}\ }\textbf {\bibinfo {volume} {96}},\
  \bibinfo {pages} {041205} (\bibinfo {year} {2017}{\natexlab{a}})}\BibitemShut
  {NoStop}%
\bibitem [{\citenamefont {Yan}\ and\ \citenamefont
  {Wang}(2017)}]{FWeyl_PhysRevB.96.041206}%
  \BibitemOpen
  \bibfield  {author} {\bibinfo {author} {\bibfnamefont {Z.}~\bibnamefont
  {Yan}}\ and\ \bibinfo {author} {\bibfnamefont {Z.}~\bibnamefont {Wang}},\
  }\href {\doibase 10.1103/PhysRevB.96.041206} {\bibfield  {journal} {\bibinfo
  {journal} {Phys. Rev. B}\ }\textbf {\bibinfo {volume} {96}},\ \bibinfo
  {pages} {041206} (\bibinfo {year} {2017})}\BibitemShut {NoStop}%
\bibitem [{\citenamefont {Bucciantini}\ \emph {et~al.}(2017)\citenamefont
  {Bucciantini}, \citenamefont {Roy}, \citenamefont {Kitamura},\ and\
  \citenamefont {Oka}}]{Fweyl_PhysRevB.96.041126}%
  \BibitemOpen
  \bibfield  {author} {\bibinfo {author} {\bibfnamefont {L.}~\bibnamefont
  {Bucciantini}}, \bibinfo {author} {\bibfnamefont {S.}~\bibnamefont {Roy}},
  \bibinfo {author} {\bibfnamefont {S.}~\bibnamefont {Kitamura}}, \ and\
  \bibinfo {author} {\bibfnamefont {T.}~\bibnamefont {Oka}},\ }\href {\doibase
  10.1103/PhysRevB.96.041126} {\bibfield  {journal} {\bibinfo  {journal} {Phys.
  Rev. B}\ }\textbf {\bibinfo {volume} {96}},\ \bibinfo {pages} {041126}
  (\bibinfo {year} {2017})}\BibitemShut {NoStop}%
\bibitem [{\citenamefont {Deb}\ and\ \citenamefont
  {Sen}(2017)}]{Fweyl_PhysRevB.95.144311}%
  \BibitemOpen
  \bibfield  {author} {\bibinfo {author} {\bibfnamefont {O.}~\bibnamefont
  {Deb}}\ and\ \bibinfo {author} {\bibfnamefont {D.}~\bibnamefont {Sen}},\
  }\href {\doibase 10.1103/PhysRevB.95.144311} {\bibfield  {journal} {\bibinfo
  {journal} {Phys. Rev. B}\ }\textbf {\bibinfo {volume} {95}},\ \bibinfo
  {pages} {144311} (\bibinfo {year} {2017})}\BibitemShut {NoStop}%
\bibitem [{\citenamefont
  {Ezawa}(2017{\natexlab{b}})}]{Fweyl2_PhysRevB.96.041205}%
  \BibitemOpen
  \bibfield  {author} {\bibinfo {author} {\bibfnamefont {M.}~\bibnamefont
  {Ezawa}},\ }\href {\doibase 10.1103/PhysRevB.96.041205} {\bibfield  {journal}
  {\bibinfo  {journal} {Phys. Rev. B}\ }\textbf {\bibinfo {volume} {96}},\
  \bibinfo {pages} {041205} (\bibinfo {year} {2017}{\natexlab{b}})}\BibitemShut
  {NoStop}%
\bibitem [{\citenamefont {Yan}\ and\ \citenamefont
  {Wang}(2016)}]{Fweyl_PhysRevLett.117.087402}%
  \BibitemOpen
  \bibfield  {author} {\bibinfo {author} {\bibfnamefont {Z.}~\bibnamefont
  {Yan}}\ and\ \bibinfo {author} {\bibfnamefont {Z.}~\bibnamefont {Wang}},\
  }\href {\doibase 10.1103/PhysRevLett.117.087402} {\bibfield  {journal}
  {\bibinfo  {journal} {Phys. Rev. Lett.}\ }\textbf {\bibinfo {volume} {117}},\
  \bibinfo {pages} {087402} (\bibinfo {year} {2016})}\BibitemShut {NoStop}%
\bibitem [{\citenamefont {Shirley}(1965)}]{PhysRev.138.B979}%
  \BibitemOpen
  \bibfield  {author} {\bibinfo {author} {\bibfnamefont {J.~H.}\ \bibnamefont
  {Shirley}},\ }\href {\doibase 10.1103/PhysRev.138.B979} {\bibfield  {journal}
  {\bibinfo  {journal} {Phys. Rev.}\ }\textbf {\bibinfo {volume} {138}},\
  \bibinfo {pages} {B979} (\bibinfo {year} {1965})}\BibitemShut {NoStop}%
\bibitem [{\citenamefont {Basov}\ \emph {et~al.}(2017)\citenamefont {Basov},
  \citenamefont {Averitt},\ and\ \citenamefont {Hsieh}}]{basov2017towards}%
  \BibitemOpen
  \bibfield  {author} {\bibinfo {author} {\bibfnamefont {D.}~\bibnamefont
  {Basov}}, \bibinfo {author} {\bibfnamefont {R.}~\bibnamefont {Averitt}}, \
  and\ \bibinfo {author} {\bibfnamefont {D.}~\bibnamefont {Hsieh}},\ }\href
  {https://www.nature.com/articles/nmat5017.pdf} {\bibfield  {journal}
  {\bibinfo  {journal} {Nature materials}\ }\textbf {\bibinfo {volume} {16}},\
  \bibinfo {pages} {1077} (\bibinfo {year} {2017})}\BibitemShut {NoStop}%
\bibitem [{\citenamefont {Oka}\ and\ \citenamefont
  {Kitamura}(2018)}]{oka2018floquet}%
  \BibitemOpen
  \bibfield  {author} {\bibinfo {author} {\bibfnamefont {T.}~\bibnamefont
  {Oka}}\ and\ \bibinfo {author} {\bibfnamefont {S.}~\bibnamefont {Kitamura}},\
  }\href {https://arxiv.org/pdf/1804.03212.pdf} {\bibfield  {journal} {\bibinfo
   {journal} {arXiv preprint arXiv:1804.03212}\ } (\bibinfo {year}
  {2018})}\BibitemShut {NoStop}%
\bibitem [{\citenamefont {Kirilyuk}\ \emph {et~al.}(2010)\citenamefont
  {Kirilyuk}, \citenamefont {Kimel},\ and\ \citenamefont
  {Rasing}}]{Pkirilyuk2010ultrafast}%
  \BibitemOpen
  \bibfield  {author} {\bibinfo {author} {\bibfnamefont {A.}~\bibnamefont
  {Kirilyuk}}, \bibinfo {author} {\bibfnamefont {A.~V.}\ \bibnamefont {Kimel}},
  \ and\ \bibinfo {author} {\bibfnamefont {T.}~\bibnamefont {Rasing}},\ }\href
  {https://journals.aps.org/rmp/abstract/10.1103/RevModPhys.82.2731} {\bibfield
   {journal} {\bibinfo  {journal} {Rev. Mod. Phys.}\ }\textbf {\bibinfo
  {volume} {82}},\ \bibinfo {pages} {2731} (\bibinfo {year}
  {2010})}\BibitemShut {NoStop}%
\bibitem [{\citenamefont {Mikhaylovskiy}\ \emph {et~al.}(2015)\citenamefont
  {Mikhaylovskiy}, \citenamefont {Hendry}, \citenamefont {Secchi},
  \citenamefont {Mentink}, \citenamefont {Eckstein}, \citenamefont {Wu},
  \citenamefont {Pisarev}, \citenamefont {Kruglyak}, \citenamefont
  {Katsnelson}, \citenamefont {Rasing} \emph
  {et~al.}}]{mikhaylovskiy2015ultrafast}%
  \BibitemOpen
  \bibfield  {author} {\bibinfo {author} {\bibfnamefont {R.}~\bibnamefont
  {Mikhaylovskiy}}, \bibinfo {author} {\bibfnamefont {E.}~\bibnamefont
  {Hendry}}, \bibinfo {author} {\bibfnamefont {A.}~\bibnamefont {Secchi}},
  \bibinfo {author} {\bibfnamefont {J.~H.}\ \bibnamefont {Mentink}}, \bibinfo
  {author} {\bibfnamefont {M.}~\bibnamefont {Eckstein}}, \bibinfo {author}
  {\bibfnamefont {A.}~\bibnamefont {Wu}}, \bibinfo {author} {\bibfnamefont
  {R.}~\bibnamefont {Pisarev}}, \bibinfo {author} {\bibfnamefont
  {V.}~\bibnamefont {Kruglyak}}, \bibinfo {author} {\bibfnamefont
  {M.}~\bibnamefont {Katsnelson}}, \bibinfo {author} {\bibfnamefont
  {T.}~\bibnamefont {Rasing}},  \emph {et~al.},\ }\href
  {https://www.nature.com/articles/ncomms9190} {\bibfield  {journal} {\bibinfo
  {journal} {Nat. Commun.}\ }\textbf {\bibinfo {volume} {6}},\ \bibinfo {pages}
  {8190} (\bibinfo {year} {2015})}\BibitemShut {NoStop}%
\bibitem [{\citenamefont {Mentink}\ \emph {et~al.}(2015)\citenamefont
  {Mentink}, \citenamefont {Balzer},\ and\ \citenamefont
  {Eckstein}}]{mentink2015ultrafast}%
  \BibitemOpen
  \bibfield  {author} {\bibinfo {author} {\bibfnamefont {J.}~\bibnamefont
  {Mentink}}, \bibinfo {author} {\bibfnamefont {K.}~\bibnamefont {Balzer}}, \
  and\ \bibinfo {author} {\bibfnamefont {M.}~\bibnamefont {Eckstein}},\ }\href
  {https://www.nature.com/articles/ncomms7708} {\bibfield  {journal} {\bibinfo
  {journal} {Nat. Commun.}\ }\textbf {\bibinfo {volume} {6}},\ \bibinfo {pages}
  {6708} (\bibinfo {year} {2015})}\BibitemShut {NoStop}%
\bibitem [{\citenamefont {Mentink}\ and\ \citenamefont
  {Eckstein}(2014)}]{Mentink2014}%
  \BibitemOpen
  \bibfield  {author} {\bibinfo {author} {\bibfnamefont {J.~H.}\ \bibnamefont
  {Mentink}}\ and\ \bibinfo {author} {\bibfnamefont {M.}~\bibnamefont
  {Eckstein}},\ }\href {\doibase 10.1103/PhysRevLett.113.057201} {\bibfield
  {journal} {\bibinfo  {journal} {Phys. Rev. Lett.}\ }\textbf {\bibinfo
  {volume} {113}} (\bibinfo {year} {2014}),\ 10.1103/PhysRevLett.113.057201},\
  \Eprint {http://arxiv.org/abs/1401.5308} {arXiv:1401.5308} \BibitemShut
  {NoStop}%
\bibitem [{\citenamefont {Mentink}(2017)}]{mentink2017manipulating}%
  \BibitemOpen
  \bibfield  {author} {\bibinfo {author} {\bibfnamefont {J.}~\bibnamefont
  {Mentink}},\ }\href
  {http://iopscience.iop.org/article/10.1088/1361-648X/aa8abf/meta} {\bibfield
  {journal} {\bibinfo  {journal} {Journal of Physics: Condensed Matter}\
  }\textbf {\bibinfo {volume} {29}},\ \bibinfo {pages} {453001} (\bibinfo
  {year} {2017})}\BibitemShut {NoStop}%
\bibitem [{\citenamefont {Bukov}\ \emph {et~al.}(2016)\citenamefont {Bukov},
  \citenamefont {Kolodrubetz},\ and\ \citenamefont {Polkovnikov}}]{Bukov2016}%
  \BibitemOpen
  \bibfield  {author} {\bibinfo {author} {\bibfnamefont {M.}~\bibnamefont
  {Bukov}}, \bibinfo {author} {\bibfnamefont {M.}~\bibnamefont {Kolodrubetz}},
  \ and\ \bibinfo {author} {\bibfnamefont {A.}~\bibnamefont {Polkovnikov}},\
  }\href {\doibase 10.1103/PhysRevLett.116.125301} {\bibfield  {journal}
  {\bibinfo  {journal} {Phys. Rev. Lett.}\ }\textbf {\bibinfo {volume} {116}},\
  \bibinfo {pages} {1} (\bibinfo {year} {2016})},\ \Eprint
  {http://arxiv.org/abs/1510.02744} {arXiv:1510.02744} \BibitemShut {NoStop}%
\bibitem [{\citenamefont {Liu}\ \emph {et~al.}(2018)\citenamefont {Liu},
  \citenamefont {Hejazi},\ and\ \citenamefont
  {Balents}}]{hejazi1PhysRevLett.121.107201}%
  \BibitemOpen
  \bibfield  {author} {\bibinfo {author} {\bibfnamefont {J.}~\bibnamefont
  {Liu}}, \bibinfo {author} {\bibfnamefont {K.}~\bibnamefont {Hejazi}}, \ and\
  \bibinfo {author} {\bibfnamefont {L.}~\bibnamefont {Balents}},\ }\href
  {\doibase 10.1103/PhysRevLett.121.107201} {\bibfield  {journal} {\bibinfo
  {journal} {Phys. Rev. Lett.}\ }\textbf {\bibinfo {volume} {121}},\ \bibinfo
  {pages} {107201} (\bibinfo {year} {2018})}\BibitemShut {NoStop}%
\bibitem [{\citenamefont {Hejazi}\ \emph {et~al.}(2019)\citenamefont {Hejazi},
  \citenamefont {Liu},\ and\ \citenamefont {Balents}}]{hejazi2018floquet}%
  \BibitemOpen
  \bibfield  {author} {\bibinfo {author} {\bibfnamefont {K.}~\bibnamefont
  {Hejazi}}, \bibinfo {author} {\bibfnamefont {J.}~\bibnamefont {Liu}}, \ and\
  \bibinfo {author} {\bibfnamefont {L.}~\bibnamefont {Balents}},\ }\href
  {\doibase 10.1103/PhysRevB.99.205111} {\bibfield  {journal} {\bibinfo
  {journal} {Phys. Rev. B}\ }\textbf {\bibinfo {volume} {99}},\ \bibinfo
  {pages} {205111} (\bibinfo {year} {2019})}\BibitemShut {NoStop}%
\bibitem [{\citenamefont {Quito}\ and\ \citenamefont
  {Flint}(2020{\natexlab{a}})}]{quito2020floquet}%
  \BibitemOpen
  \bibfield  {author} {\bibinfo {author} {\bibfnamefont {V.}~\bibnamefont
  {Quito}}\ and\ \bibinfo {author} {\bibfnamefont {R.}~\bibnamefont {Flint}},\
  }\href {https://arxiv.org/abs/2003.04272} {\bibfield  {journal} {\bibinfo
  {journal} {arXiv preprint arXiv:2003.04272}\ } (\bibinfo {year}
  {2020}{\natexlab{a}})}\BibitemShut {NoStop}%
\bibitem [{\citenamefont {Quito}\ and\ \citenamefont
  {Flint}(2020{\natexlab{b}})}]{quito2020polarization}%
  \BibitemOpen
  \bibfield  {author} {\bibinfo {author} {\bibfnamefont {V.}~\bibnamefont
  {Quito}}\ and\ \bibinfo {author} {\bibfnamefont {R.}~\bibnamefont {Flint}},\
  }\href {https://arxiv.org/abs/2003.05933} {\bibfield  {journal} {\bibinfo
  {journal} {arXiv preprint arXiv:2003.05933}\ } (\bibinfo {year}
  {2020}{\natexlab{b}})}\BibitemShut {NoStop}%
\bibitem [{\citenamefont {Chaudhary}\ \emph {et~al.}(2019)\citenamefont
  {Chaudhary}, \citenamefont {Hsieh},\ and\ \citenamefont
  {Refael}}]{PhysRevB.100.220403}%
  \BibitemOpen
  \bibfield  {author} {\bibinfo {author} {\bibfnamefont {S.}~\bibnamefont
  {Chaudhary}}, \bibinfo {author} {\bibfnamefont {D.}~\bibnamefont {Hsieh}}, \
  and\ \bibinfo {author} {\bibfnamefont {G.}~\bibnamefont {Refael}},\ }\href
  {\doibase 10.1103/PhysRevB.100.220403} {\bibfield  {journal} {\bibinfo
  {journal} {Phys. Rev. B}\ }\textbf {\bibinfo {volume} {100}},\ \bibinfo
  {pages} {220403} (\bibinfo {year} {2019})}\BibitemShut {NoStop}%
\bibitem [{\citenamefont {Ron}\ \emph {et~al.}(2019)\citenamefont {Ron},
  \citenamefont {Chaudhary}, \citenamefont {Zhang}, \citenamefont {Ning},
  \citenamefont {Zoghlin}, \citenamefont {Wilson}, \citenamefont {Averitt},
  \citenamefont {Refael},\ and\ \citenamefont {Hsieh}}]{ron2019ultrafast}%
  \BibitemOpen
  \bibfield  {author} {\bibinfo {author} {\bibfnamefont {A.}~\bibnamefont
  {Ron}}, \bibinfo {author} {\bibfnamefont {S.}~\bibnamefont {Chaudhary}},
  \bibinfo {author} {\bibfnamefont {G.}~\bibnamefont {Zhang}}, \bibinfo
  {author} {\bibfnamefont {H.}~\bibnamefont {Ning}}, \bibinfo {author}
  {\bibfnamefont {E.}~\bibnamefont {Zoghlin}}, \bibinfo {author} {\bibfnamefont
  {S.}~\bibnamefont {Wilson}}, \bibinfo {author} {\bibfnamefont
  {R.}~\bibnamefont {Averitt}}, \bibinfo {author} {\bibfnamefont
  {G.}~\bibnamefont {Refael}}, \ and\ \bibinfo {author} {\bibfnamefont
  {D.}~\bibnamefont {Hsieh}},\ }\href {https://arxiv.org/abs/1910.06376}
  {\bibfield  {journal} {\bibinfo  {journal} {arXiv preprint arXiv:1910.06376}\
  } (\bibinfo {year} {2019})}\BibitemShut {NoStop}%
\bibitem [{\citenamefont {Sivadas}\ \emph {et~al.}(2015)\citenamefont
  {Sivadas}, \citenamefont {Daniels}, \citenamefont {Swendsen}, \citenamefont
  {Okamoto},\ and\ \citenamefont {Xiao}}]{Sivadas2015}%
  \BibitemOpen
  \bibfield  {author} {\bibinfo {author} {\bibfnamefont {N.}~\bibnamefont
  {Sivadas}}, \bibinfo {author} {\bibfnamefont {M.~W.}\ \bibnamefont
  {Daniels}}, \bibinfo {author} {\bibfnamefont {R.~H.}\ \bibnamefont
  {Swendsen}}, \bibinfo {author} {\bibfnamefont {S.}~\bibnamefont {Okamoto}}, \
  and\ \bibinfo {author} {\bibfnamefont {D.}~\bibnamefont {Xiao}},\ }\href
  {\doibase 10.1103/PhysRevB.91.235425} {\bibfield  {journal} {\bibinfo
  {journal} {Phys. Rev. B}\ }\textbf {\bibinfo {volume} {91}},\ \bibinfo
  {pages} {235425} (\bibinfo {year} {2015})}\BibitemShut {NoStop}%
\bibitem [{\citenamefont {Liu}\ \emph {et~al.}()\citenamefont {Liu},
  \citenamefont {Hejazi},\ and\ \citenamefont {Balents}}]{Liu2018}%
  \BibitemOpen
  \bibfield  {author} {\bibinfo {author} {\bibfnamefont {J.}~\bibnamefont
  {Liu}}, \bibinfo {author} {\bibfnamefont {K.}~\bibnamefont {Hejazi}}, \ and\
  \bibinfo {author} {\bibfnamefont {L.}~\bibnamefont {Balents}},\ }\href
  {http://arxiv.org/abs/1801.00401} {\ }\Eprint
  {http://arxiv.org/abs/1801.00401} {arXiv:1801.00401} \BibitemShut {NoStop}%
\bibitem [{\citenamefont {Mattis}(2006)}]{mattis2006theory}%
  \BibitemOpen
  \bibfield  {author} {\bibinfo {author} {\bibfnamefont {D.~C.}\ \bibnamefont
  {Mattis}},\ }\href@noop {} {\emph {\bibinfo {title} {The theory of magnetism
  made simple: an introduction to physical concepts and to some useful
  mathematical methods}}}\ (\bibinfo  {publisher} {World Scientific Publishing
  Company},\ \bibinfo {year} {2006})\BibitemShut {NoStop}%
\bibitem [{\citenamefont {Chen}\ \emph {et~al.}(2011)\citenamefont {Chen},
  \citenamefont {Nascimb\`ene}, \citenamefont {Aidelsburger}, \citenamefont
  {Atala}, \citenamefont {Trotzky},\ and\ \citenamefont
  {Bloch}}]{PhysRevLett.107.210405}%
  \BibitemOpen
  \bibfield  {author} {\bibinfo {author} {\bibfnamefont {Y.-A.}\ \bibnamefont
  {Chen}}, \bibinfo {author} {\bibfnamefont {S.}~\bibnamefont {Nascimb\`ene}},
  \bibinfo {author} {\bibfnamefont {M.}~\bibnamefont {Aidelsburger}}, \bibinfo
  {author} {\bibfnamefont {M.}~\bibnamefont {Atala}}, \bibinfo {author}
  {\bibfnamefont {S.}~\bibnamefont {Trotzky}}, \ and\ \bibinfo {author}
  {\bibfnamefont {I.}~\bibnamefont {Bloch}},\ }\href {\doibase
  10.1103/PhysRevLett.107.210405} {\bibfield  {journal} {\bibinfo  {journal}
  {Phys. Rev. Lett.}\ }\textbf {\bibinfo {volume} {107}},\ \bibinfo {pages}
  {210405} (\bibinfo {year} {2011})}\BibitemShut {NoStop}%
\bibitem [{\citenamefont {G{\"o}rg}\ \emph {et~al.}(2018)\citenamefont
  {G{\"o}rg}, \citenamefont {Messer}, \citenamefont {Sandholzer}, \citenamefont
  {Jotzu}, \citenamefont {Desbuquois},\ and\ \citenamefont
  {Esslinger}}]{gorg2018enhancement}%
  \BibitemOpen
  \bibfield  {author} {\bibinfo {author} {\bibfnamefont {F.}~\bibnamefont
  {G{\"o}rg}}, \bibinfo {author} {\bibfnamefont {M.}~\bibnamefont {Messer}},
  \bibinfo {author} {\bibfnamefont {K.}~\bibnamefont {Sandholzer}}, \bibinfo
  {author} {\bibfnamefont {G.}~\bibnamefont {Jotzu}}, \bibinfo {author}
  {\bibfnamefont {R.}~\bibnamefont {Desbuquois}}, \ and\ \bibinfo {author}
  {\bibfnamefont {T.}~\bibnamefont {Esslinger}},\ }\href
  {https://www.nature.com/articles/nature25135} {\bibfield  {journal} {\bibinfo
   {journal} {Nature}\ }\textbf {\bibinfo {volume} {553}},\ \bibinfo {pages}
  {481} (\bibinfo {year} {2018})}\BibitemShut {NoStop}%
\bibitem [{\citenamefont {Chittari}\ \emph {et~al.}(2016)\citenamefont
  {Chittari}, \citenamefont {Park}, \citenamefont {Lee}, \citenamefont {Han},
  \citenamefont {MacDonald}, \citenamefont {Hwang},\ and\ \citenamefont
  {Jung}}]{chittari_PhysRevB.94.184428}%
  \BibitemOpen
  \bibfield  {author} {\bibinfo {author} {\bibfnamefont {B.~L.}\ \bibnamefont
  {Chittari}}, \bibinfo {author} {\bibfnamefont {Y.}~\bibnamefont {Park}},
  \bibinfo {author} {\bibfnamefont {D.}~\bibnamefont {Lee}}, \bibinfo {author}
  {\bibfnamefont {M.}~\bibnamefont {Han}}, \bibinfo {author} {\bibfnamefont
  {A.~H.}\ \bibnamefont {MacDonald}}, \bibinfo {author} {\bibfnamefont
  {E.}~\bibnamefont {Hwang}}, \ and\ \bibinfo {author} {\bibfnamefont
  {J.}~\bibnamefont {Jung}},\ }\href {\doibase 10.1103/PhysRevB.94.184428}
  {\bibfield  {journal} {\bibinfo  {journal} {Phys. Rev. B}\ }\textbf {\bibinfo
  {volume} {94}},\ \bibinfo {pages} {184428} (\bibinfo {year}
  {2016})}\BibitemShut {NoStop}%
\bibitem [{\citenamefont {Casto}\ \emph {et~al.}(2015)\citenamefont {Casto},
  \citenamefont {Clune}, \citenamefont {Yokosuk}, \citenamefont {Musfeldt},
  \citenamefont {Williams}, \citenamefont {Zhuang}, \citenamefont {Lin},
  \citenamefont {Xiao}, \citenamefont {Hennig}, \citenamefont {Sales} \emph
  {et~al.}}]{casto2015strong}%
  \BibitemOpen
  \bibfield  {author} {\bibinfo {author} {\bibfnamefont {L.}~\bibnamefont
  {Casto}}, \bibinfo {author} {\bibfnamefont {A.}~\bibnamefont {Clune}},
  \bibinfo {author} {\bibfnamefont {M.}~\bibnamefont {Yokosuk}}, \bibinfo
  {author} {\bibfnamefont {J.}~\bibnamefont {Musfeldt}}, \bibinfo {author}
  {\bibfnamefont {T.}~\bibnamefont {Williams}}, \bibinfo {author}
  {\bibfnamefont {H.}~\bibnamefont {Zhuang}}, \bibinfo {author} {\bibfnamefont
  {M.-W.}\ \bibnamefont {Lin}}, \bibinfo {author} {\bibfnamefont
  {K.}~\bibnamefont {Xiao}}, \bibinfo {author} {\bibfnamefont {R.}~\bibnamefont
  {Hennig}}, \bibinfo {author} {\bibfnamefont {B.}~\bibnamefont {Sales}},
  \emph {et~al.},\ }\href {https://aip.scitation.org/doi/abs/10.1063/1.4914134}
  {\bibfield  {journal} {\bibinfo  {journal} {APL materials}\ }\textbf
  {\bibinfo {volume} {3}},\ \bibinfo {pages} {041515} (\bibinfo {year}
  {2015})}\BibitemShut {NoStop}%
\bibitem [{\citenamefont {Bravyi}\ \emph {et~al.}(2011)\citenamefont {Bravyi},
  \citenamefont {DiVincenzo},\ and\ \citenamefont
  {Loss}}]{bravyi2011schrieffer}%
  \BibitemOpen
  \bibfield  {author} {\bibinfo {author} {\bibfnamefont {S.}~\bibnamefont
  {Bravyi}}, \bibinfo {author} {\bibfnamefont {D.~P.}\ \bibnamefont
  {DiVincenzo}}, \ and\ \bibinfo {author} {\bibfnamefont {D.}~\bibnamefont
  {Loss}},\ }\href
  {https://www.sciencedirect.com/science/article/pii/S0003491611001059}
  {\bibfield  {journal} {\bibinfo  {journal} {Annals of physics}\ }\textbf
  {\bibinfo {volume} {326}},\ \bibinfo {pages} {2793} (\bibinfo {year}
  {2011})}\BibitemShut {NoStop}%
\end{thebibliography}%
\end{document}